\documentclass[aip,jcp,reprint,showpacs,superscriptaddress]{revtex4-2}
\usepackage{amsmath,amssymb,mathrsfs,amsfonts,dsfont}
\usepackage{bm}
\usepackage{color}
\usepackage{appendix}
\usepackage{graphicx}
\usepackage{footmisc}
\usepackage{bm}
\usepackage{color}
\usepackage{CJKutf8}
\usepackage{orcidlink}
\usepackage{hyperref}
\hypersetup{colorlinks,
	citecolor=blue,
	anchorcolor=red,
	menucolor=red,
	linkcolor=blue,
	filecolor=red,
	runcolor=red,
	urlcolor=blue,
	frenchlinks=red}

\begin{document}
\begin{CJK}{UTF8}{gbsn}
  \title{Quantum decoherence dynamics in stochastically fluctuating environments}
  
   \author{Xiangji Cai~\orcidlink{0000-0001-6655-5736}}
   \email[Authors to whom correspondence should be addressed: ]{xiangjicai@foxmail.com; ygpeng@sdu.edu.cn; and yzheng@sdu.edu.cn}
   \affiliation{School of Science, Shandong Jianzhu University, Jinan 250101, China}
   \author{Yanyan Feng}
   \affiliation{School of Science, Shandong Jianzhu University, Jinan 250101, China}
   \author{Jing Ren}
   \affiliation{School of Science, Shandong Jianzhu University, Jinan 250101, China}
   \author{Yonggang Peng~\orcidlink{0000-0003-0931-3057}}
   \email[Authors to whom correspondence should be addressed: ]{xiangjicai@foxmail.com; ygpeng@sdu.edu.cn; and yzheng@sdu.edu.cn}
   \affiliation{School of Physics, Shandong University, Jinan 250100, China}
   \author{Yujun Zheng~\orcidlink{0000-0001-9974-2917}}
   \email[Authors to whom correspondence should be addressed: ]{xiangjicai@foxmail.com; ygpeng@sdu.edu.cn; and yzheng@sdu.edu.cn}
   \affiliation{School of Physics, Shandong University, Jinan 250100, China}
   
\date{\today}

\begin{abstract}
	We theoretically study the decoherence of a two-level quantum system coupled to noisy environments exhibiting linear and quadratic fluctuations within the framework of a stochastic Liouville equation. It is shown that the intrinsic energy levels of the quantum system renormalize under either the linear or quadratic influence of the environmental noise. In the case of quadratic dependence, the renormalization of the energy levels of the system emerges even if the environmental noise exhibits stationary statistical properties. This is in contrast to the case under linear influence, where the intrinsic energy levels of the system renormalize only if the environmental noise displays nonstationary statistics. We derive the analytical expressions of the decoherence function in the cases where the fluctuation of the frequency difference depends linearly and quadratically on the nonstationary Ornstein–Uhlenbeck noise (OUN) and random telegraph noise (RTN) processes, respectively. In the case of the linear dependence of the OUN, the environmental nonstationary statistical property can enhance the dynamical decoherence. However, the nonstationary statistics of the environmental noise can suppress the quantum decoherence in this case under the quadratic influence of the OUN. In the presence of the RTN, the quadratic influence of the environmental noise does not give rise to decoherence but only causes a determinate frequency renormalization in dynamical evolution. The environmental nonstationary statistical property can suppress the quantum decoherence of the case under the linear influence of the RTN.
\end{abstract}
\maketitle
\section{Introduction}
\label{sec:intr}

Quantum coherence, as a central feature in microscopic systems, increasingly enables many applications that are impossible to realize in classical mechanics and ray optics~\cite{Breuerbook,RevModPhys.89.041003,PhysRep.762-764.1}.
Any realistic quantum system is never completely isolated but always interacts with its surrounding environments, and the unavoidable interaction with the surroundings destroys the coherence between the states of an open quantum system in its dynamical evolution.
The destruction of the coherence of quantum systems induced by the environments is the main impediment to realizing experimental devices for quantum information processing. 
Many efforts have been devoted to finding physical systems with intrinsically long coherence times and realizing coherent manipulation of open quantum systems, which has a wide range of applications in quantum information science including quantum computation, quantum communication, quantum simulation and so on~\cite{RevModPhys.73.565,RevModPhys.75.457,RevModPhys.75.715,RevModPhys.86.153,Schlosshauerbook,PhysRep.831.1}. 

It has been observed experimentally the effects of dynamical decoherence in realistic physical systems, owing to the rapid advances in experimental techniques~\cite{RevModPhys.73.357,RevModPhys.75.281,PhysRep.469.155,RevModPhys.84.157}.
For example, the dynamical decoherence processes dominated by low-frequency environmental fluctuations were observed experimentally in most quantum computing nanodevices, such as, flux, charge and phase superconducting qubits~\cite{RevModPhys.86.153,RevModPhys.95.025003}.
The investigations on the decoherence dynamics of open quantum systems have drawn increasing attention, which aims at searching for the dynamical description of the quantum-to-classical transition and the mechanism of quantum decoherence in different kinds of environments~\cite{PhysRevA58.2733,PhysRevB77.174509,PhysRevA81.040103,PhysRevA102.062423,PhysRevA105.052443,PhysRevE72.056106,PhysRevA71.020101}.
Due to the difficulty in dealing with higher-order environmental correlations, the quantum dynamics can be solved exactly in a few physical models~\cite{PhysRevA81.042103,PhysRevA81.062120,PhysRevA90.062104,PhysRevLett.116.120402,PhysRevLett.109.170402,PhysRevRes.4.033151,JChemPhys.148.164101,JChemPhys.148.174102,JChemPhys.156.134114}. 
In most cases, we can only obtain the quantum dynamics of an open system under the Markovian approximations by casting away the environmental correlations of higher-orders.
However, in some important physical situations, the dynamical evolution of the quantum system displays a memory effect and these approximations are no longer valid.
The higher-order environmental correlations gives rise to some novel non-Markovian behavior in quantum dynamics~\cite{PhysRevA59.1633,PhysRevA60.91,PhysRevLett.100.180402,PhysRevLett.105.240403,PhysRevLett.117.230401,JChemPhys.151.044102,RepProgPhys.77.094001,RevModPhys.88.021002,RevModPhys.89.015001}. 
Therefore, it is extremely meaningful to develop some approaches to obtain the quantum dynamics induced by the environments exactly.
This will play a fundamental role in understanding the precise dynamical evolution and has potential applications in suppressing the quantum decoherence of open systems.

Based on the classical and quantum descriptions of environmental effects, a number of theoretical frameworks have been established to study the dynamical decoherence of open quantum systems~\cite{RevModPhys.88.021002,RevModPhys.89.015001,RevModPhys.86.361,RepProgPhys.80.016001,RepProgPhys.80.056002,JPhysSocJpn.75.082001}.
Based on the phenomenological description in the classical treatment proposed by Anderson and Kubo, the environmental effects give rise to stochastic fluctuations in some observables of the quantum system~\cite{JPhysSocJpn.9.316,JPhysSocJpn.9.935}.
The statistical properties of the stochastic fluctuations characterize the environmental state and the interaction between the system and the environment in terms of the noise representation.
The Ornstein-Uhlenbeck noise (OUN) as an important Gaussian process has been widely used to study the dynamics of open quantum systems under the conditions that the environmental coupling is weak and the environment has a large number of degrees of freedom~\cite{PhysRevLett.105.015301,PhysRevLett.109.130401,PhysRevLett.105.015301,PhysRevLett.109.130401,PhysRevX9.021009,PhysRevB101.174302,PhysRevA105.L010601}. 
In some situations where the coupling between the system and the environment is not weak and the number of environmental degrees of freedom is relatively small, random telegraph noise (RTN) with non-Gaussian statistics becomes dominant in the dynamical evolution of quantum  systems~\cite{NewJPhys.8.1,PhysRevA73.022332,PhysRevB75.054515,JChemPhys.144.024113,PhysRevA100.052104,PhysRevA107.L030601,ApplPhysLett.122.244001}.

Many theoretical investigations of the dynamical decoherence in fluctuating environments primarily concentrate on the case that the fluctuation in some system observable is directly dependent on the environmental noise with stationary statistics~\cite{PhysRevLett.96.097009,PhysRevB79.125317,PhysRevA85.052125,PhysRevA89.012114,SciRep.10.88}.
These investigations effectively describe the quantum decoherence induced by the environments in most dynamical processes and physical situations.
However, in some other special dynamical processes, the fluctuation term of the system observable is an indirect function of the environmental noise which may lead to many novel physical phenomena.
For example, the fluctuation of the observable in open quantum systems displays nonlinear dependence of the environmental noise in the decoherence dynamics of solid state qubits at optimal operation points~\cite{PhysRevLett.92.178301,PhysRevLett.93.237401,PhysRevB72.134519}, in single molecule kinetics associated with chemical rate reaction and fluorophore blinking processes~\cite{JChemPhys.97.3587,PhysRevLett.74.4317,JChemPhys.121.3238} and in dissipation dynamics in the presence of complex environments~\cite{PhysRevB98.115131,JChemPhys.148.014104,JChemPhys.148.114103,JChemPhys.155.174111}.
On the other hand, the statistical properties of the environmental noise are not always stationary and the nonstationary stochastic processes have been found in a wide range of physical, chemical, and biological systems~\cite{Nature271.431,PhysRevLett.115.080602,PhysRevLett.115.080603}.
Recently, there is growing interest in the study of the dynamics of open quantum systems in the presence of nonstationary environmental fluctuations due to their significant role in some important physical situations~\cite{PhysRevB72.035328,PhysRevLett.111.250404,JChemPhys.133.241101,JChemPhys.139.024109,PhysRevA87.032338,PhysRevA94.042110,JChemPhys.149.094107,Entropy25.634,NewJPhys.22.033039,PhysRevA104.042417,PhysRevA108.062208,PRXQuantum3.020321,PhysRevRes.4.013230}.
This motivates us to study the quantum decoherence in the case that the system observable depends nonlinearly on the environmental noise with nonstationary statistical property and make a comparison on the mechanism for dynamical decoherence with the case that the system observable fluctuates linearly in terms of the environmental noise exhibiting stationary statistics.

In this paper, we focus on how the effects of the nonlinear dependence and the nonstationary statistical property of the environmental noise affect the quantum decoherence of a two-level system.
We aim to explore the mechanism for dynamical decoherence and the novel physical phenomena in quantum dynamics induced by the linear and quadratic influences of environmental noise.
In contrast to the linear case,  the quadratic dependence of the environmental noise gives rise to the  renormalization of the intrinsic energy levels of the quantum system not only when the environmental fluctuations exhibits nonstationary statistics but also in the case of the stationary statistical property of the environmental fluctuations.
Another goal of this paper is to analyze the influence of the nonstationary statistics of the environmental noise on the enhancement and suppression of the dynamical decoherence in the presence of some important environmental noise.
By means of the approach of the stochastic Liouville equation, we derive the analytical expressions of the decoherence function under the linear and quadratic influences of nonstationary OUN and RTN, respectively.
In the presence of the OUN, the environmental nonstationary statistical property suppresses and enhances the dynamical decoherence for the cases under linear and quadratic  influences of the fluctuation in the frequency difference, respectively.
In the presence of the RTN, the nonstationary statistics of the environmental noise suppress the decoherence in dynamical evolution for the case that the frequency difference fluctuates linearly, whereas it does not affect the dynamical decoherence due to the fact that the stochasticity of the fluctuation in the frequency difference vanishes for the case under quadratic  dependence of the environmental noise.

This paper is organized as follows. In Sec.~\ref{sec:theo} we introduce the theoretical framework of stochastic Liouville equation in the phenomenological description of the influence of fluctuating environments on a physical system.
In Sec.~\ref{sec:app} we apply the approach of the stochastic Liouville equation to study the quantum decoherence of  a two-level system induced by stochastic Markovian environmental fluctuations.
We derive the explicit expressions of the decoherence function when the frequency difference of the system fluctuates linearly and quadratically in  terms of nonstationary OUN or RTN, respectively.
We analyze the effects of the nonlinear fluctuation in the frequency difference and the nonstationary statistical property of the environmental noise on the decoherence dynamics in detail.
In Sec.~\ref{sec:con} we give the conclusions from the present study.

\section{Theoretical framework of stochastic Liouville equation}
\label{sec:theo}

Due to the omnipresence of environmental noise, the dynamical characteristics of a physical system are closely dependent on the environments. 
Investigations of the environmental effects on the system arise in numerous situations and involve many different approaches.
The environmental influence on the physical properties of a system can be generally described at the level of a phenomenological equation via some external parameters based on stochastic theory~\cite{JPhysSocJpn.9.316,JPhysSocJpn.9.935}. 
If the system is coupled to a static environment, these parameters are turned into stochastic variables with definite single-probability distributions, whereas they become stochastic processes with infinite hierarchies of multi-probability distributions when the system interacts with a fluctuating environment~\cite{Kampenbook}.

In the phenomenological description, the time evolution of some physical quantity $q$ of the system under the influence of environmental fluctuations satisfies a stochastic differential equation of the form~\cite{Kampenbook}
\begin{equation}
	\label{eq:stoevo}
	\frac{d}{dt}q\bm{(}t;\{x(t)\}\bm{)}=f[q\bm{(}t;\{x(t)\}\bm{)},t;x(t)],
\end{equation}
where $q\bm{(}t;\{x(t)\}\bm{)}$ and $f[q\bm{(}t;\{x(t)\}\bm{)},t;x(t)]$ are the state variable and functional relation in terms of the external noise $x(t)$, respectively.
The environmental noise $x(t)$ is subject to a classical stochastic process and its statistical properties characterize the state of the environment and its interaction with the system.
Generally, $q$ and $f$ can be $n$-dimensional vectors and then Eq.~\eqref{eq:stoevo} corresponds to a system of $n$ coupled stochastic differential equations.
The environmental noise $x(t)$ can also be subject to a composite stochastic process, namely, $x(t)=f\bm(x_{1}(t),\cdots,x_{n}(t)\bm)$ as a function of some other stochastic processes.

Based on the statistical theory, we actually observe the ensemble-average quantity $q(t)$ rather than the stochastic quantity $q\bm{(}t;x(t)\bm{)}$ directly.
The physical quantity $q(t)$, being a statistical average over all realizations of environmental fluctuations, can be formally expressed as 
\begin{equation}
	\label{eq:avequa}
	q(t)=\langle q\bm{(}t;\{x(t)\}\bm{)}\rangle=\int q\bm{(}t;\{x(t)\}\bm{)}P[x(t)]dx(t),
\end{equation}
where $\langle\cdots\rangle$ denotes the ensemble-average taken over the environmental noise $x(t)$ and $P[x(t)]$ is the probability distribution of different realizations of $x(t)$.
In contrast to the case in a static environment, the statistical average quantity $q(t)$ in Eq.~\eqref{eq:avequa} involves no longer just a definite single-point probability $P(x, t)$ but an infinite hierarchy of multi-point probabilities $P(x_{1},t_{1}; \cdots; x_{n},t_{n})$.
Therefore, it is hardly possible to obtain the statistical average quantity $q(t)$ directly based on Eq.~\eqref{eq:avequa} in a fluctuating environment.

Markovian stochastic processes are of fundamental importance in a variety of research fields and have wide applications, in particular to the description of  environmental fluctuations~\cite{Kampenbook}. 
For a Markovian stochastic process $x(t)$, the whole hierarchy of multi-point probabilities $P(x_{1},t_{1};\cdots;x_{n},t_{n})$ is uniquely determined by the single-point probability $P(x,t)$ that the process $x(t)$ has the value $x$ at time $t$ and the transition probability $P(x,t|x',t')$ that the process $x(t)$ takes the value $x$ at time $t$ given that it took the value $x'$ at time $t'$ ($t'<t$).
In general, the transition probability $P(x,t|x',t')$ obeys a master equation as
\begin{equation}
	\label{eq:trapro}
	\frac{\partial}{\partial t}P(x,t|x',t')=\mathcal{M}_{x}P(x,t|x',t'),
\end{equation}
where the forward generator $\mathcal{M}_{x}$ is a differential operator only involving the derivatives with respect to $x$.

In the presence of environmental noise with Markovian statistical property, we can obtain the ensemble-average quantity $q(t)$ within the framework of the stochastic Liouville equation.
In view of the fact that the quantity $q\bm{(}t;x(t)\bm{)}$ in Eq.~\eqref{eq:stoevo} is also stochastic, we can introduce the bivariate composite process $\{q,x\}$ 
and define the joint probability as~\cite{Kampenbook,Riskenbook}
\begin{equation}
	\label{eq:joipro}
	\begin{split}
		P(q,x,t)&=\left\langle\delta\bm(q\bm{(}t;\{x(t)\}\bm{)}-q\bm)\delta\bm(x(t)-x\bm)\right\rangle\\
		&=\left\langle\delta\bm(q(x,t)-q\bm)\right\rangle|_{x(t)=x}\left\langle\delta\bm(x(t)-x\bm)\right\rangle,
	\end{split}
\end{equation}
which is the probability that the stochastic quantity $q\bm{(}t;\{x(t)\}\bm{)}$ has the value $q$ at time $t$ conditioned on the environmental noise $x(t)$ takes the value $x$.
Because the quantity $q\bm{(}t;x(t)\bm{)}$ is a function of $x(t)$, the joint probability for the composite process in Eq.~\eqref{eq:joipro} varies in time due to the flow in $q$-space which is conditional on $x(t)$ having the value $x$ at time $t$.
By combining the continuity equation for the probability of $q$ with the master equation~\eqref{eq:trapro} for the environmental noise $x(t)$, we obtain the time evolution of the joint probability $P(q,x,t)$ which describes a flow in $(q,x)$ space as
\begin{equation}
	\label{eq:evojoipro}
	\begin{split}
		\frac{\partial}{\partial t}P(q,x,t)=&-\frac{\partial}{\partial q}[f(q,x,t)P(q,x,t)]+\mathcal{M}_{x}P(q,x,t),
	\end{split}
\end{equation}
where the initial condition satisfies 
\begin{equation}
	\label{eq:iniconjoipro}
	\begin{split}
		P(q,x,0)=\delta\bm(q-q_{0}\bm)\delta(x-x_{0}),
	\end{split}
\end{equation}
with $q_{0}=q(0)=\langle q\bm{(}0;\{x(0)\}\bm{)}\rangle=q\bm{(}0;\{x(0)\}\bm{)}$.
It has been assumed that there is no initial correlation between the system and its environment, namely, the probability distributions of $q$ and $x$ are statistically independent of each other initially.
The bivariate process $\{q,x\}$ is also Markovian and the solution of Eq.~\eqref{eq:evojoipro} describes the transition probability from the state $(q_{0},x_{0})$ at the initial time $t=0$ to the final state  $(q,x)$ at time $t$.

Generally, the joint probability $P(q,x,t)$ in Eq.~\eqref{eq:evojoipro} can be solved only in rare cases.
In the special case that $f$ in Eq.~\eqref{eq:stoevo} is linear in $q$ and does not explicitly depend on time, namely, $(d/dt)q\bm{(}t;\{x(t)\}\bm{)}=f[x(t)]q\bm{(}t;\{x(t)\}\bm{)}$, we can define the marginal average over $q$ for fixed $x$ as
\begin{equation}
	\label{eq:parave}
	q(x,t)=\int qP(q,x,t)dq.
\end{equation}
By multiplying Eq.~\eqref{eq:evojoipro} with $q$ and then integrating over $q$, we can obtain the master equation for the marginal average $q(x,t)$ as
\begin{equation}
	\label{eq:stoLioequ}
	\begin{split}
		\frac{\partial}{\partial t}q(x,t)=&f(x)q(x,t)+\mathcal{M}_{x}q(x,t),
	\end{split}
\end{equation}
where the initial condition satisfies $q(x,0)=q(0)P(x,0)$.
The master equation for the marginal average in Eq.~\eqref{eq:stoLioequ} is usually called the stochastic Liouville equation.
Consequently, the ensemble-average quantity $q(t)$ can be obtained by the complete average
\begin{equation}
	\label{eq:comave}
	q(t)=\int q(x,t)dx.
\end{equation}
This is the general theoretical framework of the approach of the stochastic Liouville equation.
Therefore, by means of the marginal and complete averages in terms of Eqs.~\eqref{eq:stoLioequ} and~\eqref{eq:comave}, we can obtain the same ensemble-average quantity $q(t)$ as that expressed in Eq.~\eqref{eq:avequa}.
The equivalence between them is completely due to the Markovian statistical property of the environmental noise $x(t)$. 
It is worth noting that although we here consider the case that the environmental noise $x(t)$ takes the state $x$ with a continuous distribution, it can be directly generalized to the case of the environmental noise $x(t)$ with a set of discrete states $x_{n}$.
For the case where the environmental noise $x(t)$ takes discrete states, we just need to change the integral over $x$ in Eq.~\eqref{eq:comave} to the summation of $x_{n}$ as in the previous work~\cite{PhysRevA105.052443}.
It is also worth mentioning that the approach is not only valid for the case in the presence of environmental noise with stationary statistical properties but also for the case under the influence of environmental noise with nonstationary statistics without any approximations.

The main advantage of the approach of the stochastic Liouville equation is that we can obtain the statistical average quantity $q(t)$ numerically or even analytically without involving any environmental correlations.
However, whether we can obtain the analytical expression of the statistical average quantity $q(t)$ is contingent on both the special type of the functional relation $f[x(t)]$ and the solution of the stochastic Liouville equation~\eqref{eq:stoLioequ}.
The approach of the stochastic Liouville equation can be extended to the case in the presence of quantum noise in some special physical situations where some system observable is governed by a quantum Fokker-Planck equation rather than a master equation for the transition probability as in Eq.~\eqref{eq:trapro}~\cite{JPhysSocJpn.58.101,PhysRevA43.4131,JPhysSocJpn.75.082001}.

\section{Applications: Quantum decoherence in fluctuating environments}
\label{sec:app}

As applications, we study the decoherence of a two-level quantum system interacting with stochastically fluctuating environments that exhibit Markovian statistical properties by means of the approach of the stochastic Liouville equation. 
We consider the physical situations relevant to some experiments, where the quantum decoherence of the two-level system is dominated by pure dephasing rather than energy dissipation~\cite{PhysRevLett.88.047901,PhysRevLett.95.257002,PhysRevLett.97.167001,PhysRevLett.98.047004,PhysRevA90.042307}.
Due to the influences of a dephasing environment, the Hamiltonian of the two-level quantum system we consider here fluctuates stochastically as~(see Appendix~\ref{sec:App0})~\cite{PhysRevB77.174509,RevModPhys.86.361}
\begin{equation}
	\label{eq:stoHam}
	H(t)=\frac{\hbar}{2}[\omega_{0}+\delta\omega(t)]\sigma_{z},
\end{equation}
where $\sigma_{z}$ is the Pauli matrix, $\omega_{0}$ denotes the intrinsic frequency difference between the states $|1\rangle$ and $|0\rangle$, and the fluctuation term $\delta\omega(t)=f[x(t)]$ is a function of the environmental noise $x(t)$.
The environmental noise $x(t)$ is subject to a Markovian stochastic process and its transition probability satisfies the master equation~\eqref{eq:trapro}.

This physical description in Eq.~\eqref{eq:stoHam} is known as the spectral diffusion model which has been widely used to study linewidth broadening in single molecule spectroscopy~\cite{PhysRevLett.87.207403,PhysRevLett.90.238305,RepProgPhys.68.1129} and quantum decoherence in fluctuating environments~\cite{RevModPhys.86.361,RepProgPhys.80.016001}.
Physically, this model can be realized by spin and pseudo-spin qubit systems coupled to stochastic magnetic fields in experiments~\cite{Nature458.996,NatCommun.3.858,NatCommun.10.3715,PhysRevLett.121.023601}.
It has demonstrated the equivalence in the dephasing dynamics of the reduced system between the classical and quantum treatments of environmental effects~\cite{PhysRevA82.022111,JChemPhys.151.014109,JChemPhys.153.154115,JPhysCondensMat.29.333001,SciRep.10.22189,PhysRevA104.022202,PhysRevA104.052431}.
It is worth noting that the fluctuation term $\delta\omega(t)$ is no longer a function of a stochastic process but a time-dependent operator of the environment in the presence of quantum noise~\cite{Gardinerbook3,PhysRevLett.106.217205,PhysRevLett.116.150503,JPhysCondensMat.29.333001}.

For the initial preparation in coherent superposition of the states $|0\rangle$ and $|1\rangle$, the time evolution of the coherence of the system, namely, the off-diagonal element of the reduced density matrix $\rho_{01}(t)$, can be expressed as
\begin{equation}
	\label{eq:cohevo}
	\rho_{01}(t)=F(t)e^{i\omega_{0}t}\rho_{01}(0),
\end{equation}
where $F(t)=\left\langle\exp\left[{i\int_{0}^{t}\delta\omega(t')dt'}\right]\right\rangle=\left\langle F\bm{(}t;\{x(t)\}\bm{)}\right\rangle$ denotes the decoherence function quantifying the environmental influences on the coherence evolution of the system with the stochastic coherence function $F\bm{(}t;\{x(t)\}\bm{)}$ expressed in terms of the environmental noise $x(t)$ as
\begin{equation}
	\label{eq:decfun}
	F\bm{(}t;\{x(t)\}\bm{)}=\exp\left[{i\int_{0}^{t}f\left[x(t')\right]dt'}\right].
\end{equation}
For some special cases, the decoherence function $F(t)$ in Eq.~\eqref{eq:cohevo} is a time-dependent complex function and the dynamics of the quantum system is governed by a time-local master equation~\cite{QuantumInfProcess.5.503,PhysRevA95.052104,JChemPhys.149.094107}
\begin{equation}
	\label{eq:masequ}
	\frac{d}{dt}\rho(t)=-\frac{i}{2}[\omega_{0}-S(t)][\sigma_{z},\rho(t)]
	+\frac{1}{2}\Gamma(t)[\sigma_{z}\rho(t)\sigma_{z}-\rho(t)],
\end{equation}
where the decoherence rate $\Gamma(t)$ and the frequency shift $S(t)$ are, respectively, defined by
\begin{equation}
	\label{eq:frshdera}
	\begin{split}
		\Gamma(t)=-\mathrm{Re}\left[\frac{dF(t)/dt}{F(t)}\right],\:		S(t)=-\mathrm{Im}\left[\frac{dF(t)/dt}{F(t)}\right].
	\end{split}
\end{equation}
The decoherence rate $\Gamma(t)$ quantifies the environmental effects on the exchange of information between the system and environment while the frequency shift $S(t)$ characterizes the environmental effects on the renormalization of the intrinsic energy levels of the system.

Previously, investigations of quantum decoherence in fluctuating environments mainly focused on the case that the fluctuation in the frequency difference of the system is directly governed by the environmental noise, namely, $\delta\omega(t)=x(t)$.
In some important physical situations, however, the frequency difference of the system may be indirectly governed by the environmental noise. 
For example, the fluctuation in the frequency difference displays a quadratic dependence on the environmental noise in some dynamical decoherence and dissipation processes~\cite{PhysRevLett.92.178301,PhysRevLett.93.237401,PhysRevB72.134519,JChemPhys.148.014104,JChemPhys.148.114103,JChemPhys.155.174111}.
In recent decades, there has been increasing interest in the statistical properties of nonstationary processes~\cite{PhysRevLett.115.080602,PhysRevLett.115.080603} and quantum noise spectroscopy in the presence of nonstationary noise sources has been implemented, which has potential applications in quantum optimal control and the design of quantum gates~\cite{PRXQuantum2.030315}.
Many efforts have been devoted to the study of the dynamics of open quantum systems induced by nonstationary environmental fluctuations~\cite{PhysRevLett.111.250404,JChemPhys.133.241101,JChemPhys.139.024109,PhysRevA87.032338,PhysRevA94.042110,JChemPhys.149.094107,Entropy25.634,NewJPhys.22.033039,PhysRevA104.042417,PhysRevA108.062208,PRXQuantum3.020321,PhysRevRes.4.013230}.

In the following, we will derive the decoherence function, respectively, in the case that the frequency difference of the system is in linear and quadratic dependence of the environmental noise, namely, 
\begin{equation}
	\label{eq:flulinqua}
	\delta\omega(t)=cx^{k}(t),
\end{equation}
with the constant $c\neq0$ and the integer $k=1$ and $k=2$, respectively.
For the case where $\delta\omega(t)$ is the identity function of $x(t)$, namely, $c=1$ and $k=1$, the fluctuation term is directly affected by the environmental noise and the results will be consistent with many previous investigations.
For the quadratic case $k=2$, the fluctuation in the frequency difference of the system displays a nonlinear effect of the dependence on environmental noise.
For instance, the frequency difference of a Josephson-junction superconducting qubit usually fluctuates quadratically in terms of the magnetic flux noise at the intrinsic optimal working point~\cite{PhysRevLett.97.167001,PhysRevLett.98.047004,PhysRevA90.042307}.

We can check that the time evolution of the stochastic coherence function $F\bm{(}t;\{x(t)\}\bm{)}$ in Eq.~\eqref{eq:decfun} yields a linear differential equation as
\begin{equation}
	\label{eq:stodecevo}
	\frac{d}{dt}F\bm{(}t;\{x(t)\}\bm{)}=icx^{k}(t)F\bm{(}t;\{x(t)\}\bm{)}.
\end{equation}
By taking an average over the environmental noise $x(t)$, we can obtain the decoherence function $F(t)$ formally expressed as
\begin{equation}
	\label{eq:dyson}
	\begin{split}
		F(t)&=\left\langle\exp\left[{i\int_{0}^{t}\delta\omega(t')dt'}\right]\right\rangle\\
		&=1+\sum_{n=1}^{\infty}\frac{i^{n}}{n!}\int_{0}^{t}dt_{1}\cdots
		\int_{0}^{t}dt_{n}C_{n}(t_{1},\cdots,t_{n}),
	\end{split}
\end{equation}
with the $n$th-order cumulants of the fluctuation term $\delta\omega(t)$
\begin{equation}
	\label{eq:nthmon}
	\begin{split}
		C_{n}(t_{1},\cdots,t_{n})&=\langle\delta\omega(t_{1})\cdots\delta\omega(t_{n})\rangle_{c}\\
		&=c^{n}\langle x^{k}(t_{1})\cdots x^{k}(t_{n})\rangle_{c}.
	\end{split}
\end{equation}
The expression of the decoherence function in Eq.~\eqref{eq:dyson} in the expansion in terms of the cumulants is equivalent to that in terms of the moments of $\delta\omega(t)$, which is due to the fact that the fluctuation term is a function of a stochastic process~\cite{PhysRevA94.042110,JChemPhys.149.094107}.
If $\delta\omega(t)$ obeys Gaussian statistics, its cumulants higher than second-order vanish.
It is worth further mentioning that in the presence of quantum noise, the fluctuation term $\delta\omega(t)$ is a time-dependent operator of the environment.
For the case that $\delta\omega(t)$ at different times does not commute with each other, the decoherence function can be expressed in a time-ordering Dyson expansion in terms of the moments of the fluctuation term. 
In this case, there will be additional crossing terms if we express the decoherence function in the cumulants expansion as in Eq.~\eqref{eq:dyson} even if the fluctuation term $\delta\omega(t)$ satisfies Gaussian statistics~\cite{Physica74.215,*Physica74.239,JStatPhys.51.691}.

To identify the linear and non-linear influences of the environmental noise on the quantum decoherence of the system, we here express the decoherence function in Eq.~\eqref{eq:dyson} in cumulants expansion rather than in moments expansion due to the fact that the zero average does not always lead to odd-order moments vanishing for a general stationary noise process.
In the presence of environmental noise $x(t)$ with nonstationary statistics, the odd-order cumulants in Eq.~\eqref{eq:nthmon} always exist for both the linear ($k=1$) and quadratic ($k=2$) cases. Therefore, $F(t)$ is a time-dependent complex function and there is a time dependent frequency shift $S(t)$.
Under the influence of environmental noise exhibiting stationary statistics with a zero average $\left\langle x(t)\right\rangle=0$, the odd-order cumulants in Eq.~\eqref{eq:nthmon} vanish for the linear case whereas they still exist for the quadratic case, for instance, the average of the fluctuation term $C_{1}(t)=\left\langle \delta\omega(t)\right\rangle=c\left\langle x^{2}(t)\right\rangle\neq0$.
Therefore, $F(t)$ becomes a real function of time and there is no frequency shift in the linear case.
By contrast, $F(t)$ is still a time-dependent complex function, leading to frequency renormalization in the time evolution for the quadratic case.
This is the most important physical characteristic that distinguishes the dynamical decoherence in the case that the frequency difference of the system fluctuates quadratically from the case of linear fluctuations in the frequency difference.

In the case of the fluctuation directly related to the environmental noise, namely,  $\delta\omega(t)=x(t)$, some approaches  have been established to derive the expressions of the decoherence function exactly or approximately based on the statistical properties of the environmental noise,  such as, the partial and ordered cumulants expansions, the closed higher-order time differential equations and so on~\cite{JChemPhys.133.241101,JChemPhys.139.024109,PhysRevA87.032338,PhysRevA94.042110,JChemPhys.149.094107}.
In the general case, especially the quadratic case, we can not derive the analytical expression of the decoherence function directly based on Eq.~\eqref{eq:dyson} even if we obtain all the cumulants in Eq.~\eqref{eq:nthmon}. Sometimes, we can just derive the decoherence function approximately by truncating the environmental correlations to some finite order.
However,  by means of the stochastic Liouville equation, we can obtain the decoherence function $F(t)$ numerically or even analytically  without any approximations.

Based on the theoretical framework of the stochastic Liouville equation established in Sec.~\ref{sec:theo}, the marginal average $F(x,t)$ for the decoherence function satisfies a master equation as
\begin{equation}
	\label{eq:masequparave}
	\frac{\partial}{\partial t}F(x,t)=icx^{k}F(x,t)+\mathcal{M}_{x}F(x,t),
\end{equation}
where the initial condition fulfills $F(x,0)=F(0)P(x,0)$, with $F(0)=1$.
By further integrating the marginal average over $x$, the decoherence function $F(t)$ can be expressed as
\begin{equation}
	\label{eq:comavedecfun}
	F(t)=\int F(x,t)dx.
\end{equation}
Based on Eqs.~\eqref{eq:masequparave} and~\eqref{eq:comavedecfun}, we can obtain the decoherence function numerically or analytically which does not involve 
any environmental correlations compared to the approaches in terms of the cumulant expansions~\cite{PhysRevA85.052125} or the closed time differential equations~\cite{PhysRevA94.042110,JChemPhys.149.094107,Entropy25.634}.

Stationary stochastic processes have extensive applications, in particular for describing equilibrium environmental fluctuations in both statistical and quantum  physics~\cite{Kampenbook}.
In some important dynamical processes, the environmental fluctuations are out of equilibrium and the environmental noise exhibits nonstationary statistics.
The stochastic processes with nonstationary statistical properties have been used to describe nonequilibrium environmental fluctuations, which have practical applications in the study of the dynamics of open quantum systems in nonstationary noisy environments~\cite{PhysRevLett.111.250404,JChemPhys.133.241101,JChemPhys.139.024109,PhysRevA87.032338,PhysRevA94.042110,JChemPhys.149.094107,Entropy25.634,NewJPhys.22.033039,PhysRevA104.042417,PhysRevA108.062208,PRXQuantum3.020321,PhysRevRes.4.013230}.

In the following, we will use the approach of stochastic Liouville equation to study the quantum decoherence in the presence of environmental noise with nonstationary statistics subject to two types of important stochastic processes, namely, the OUN and RTN processes, respectively.
The statistical characteristics of a nonstationary stochastic process can be extracted from those of the standard one with stationary statistics by setting an initial nonstationary distribution~(see Appendix~\ref{sec:AppA}).
The nonstationary distribution of the environmental noise indicates physically that the environment is in a nonstationary state initially~\cite{Kampenbook}.
We will derive the decoherence function analytically in cases where the fluctuation of the frequency difference depends linearly and quadratically on the environmental noise.
We will explore the nonstationary statistical property of the environmental noise on the dynamical decoherence and analyze the differences in the decoherence dynamics between the cases under linear and quadratic  environmental fluctuations.

\subsection{Decoherence under the influence of environmental noise subject to nonstationary OUN}
The standard OUN is an important stationary Gaussian process that has been widely used to study some important issues related to the dynamics of open quantum systems~\cite{PhysRevLett.105.015301,PhysRevLett.109.130401,PhysRevLett.105.015301,PhysRevLett.109.130401,PhysRevX9.021009,PhysRevB101.174302,PhysRevA105.L010601}. 

We first consider the case that the environmental noise $x(t)$ is subject to the OUN with nonstationary statistical properties.
For the nonstationary OUN, the master equation for the transition probability can be expressed as
\begin{equation}
	\label{eq:traproOUN}
	\frac{\partial}{\partial t}P(x,t|x',t')=\gamma\frac{\partial}{\partial x}\left(x+\sigma^{2}\frac{\partial}{\partial x}\right)P(x,t|x',t'),
\end{equation}
where the decay rate $\gamma$ defines the spectral width of the environmental coupling, the standard deviation $\sigma$ of the distribution describes the strength of the environmental coupling and the initial condition satisfies $P(x,t'|x',t')=\delta(x-x')$.
For the environment in an initial nonstationary state with OUN statistics, the nonstationary statistical property of the environmental noise is determined by the single-point probability initially
\begin{equation}
	\label{eq:inisinpoiOUN}
	P(x,0)=\frac{1}{\sqrt{2\pi\sigma^{2}(1-b^{2})}}
	\exp\left[-\frac{(x-b\chi)^{2}}{2\sigma^{2}(1-b^{2})}\right],
\end{equation}
with the nonstationary parameter $|b|<1$ and the initial parameter $\chi$.
For the case $b=0$, the OUN recovers the standard one with stationary statistics.
For the case $b\neq0$ and $\chi=0$,  the OUN is also stationary but with a different variance $\sigma^{2}(1-b^{2})$ compared to the standard one.

In the presence of environmental noise governed by the nonstationary OUN, the master equation of the marginal average for the decoherence function in Eq.~\eqref{eq:masequparave} can be therefore rewritten as 
\begin{equation}
	\label{eq:maraveOUN}
	\frac{\partial}{\partial t}F(x,t)=icx^{k}F(x,t)+\gamma\frac{\partial}{\partial x}\left(x+\sigma^{2}\frac{\partial}{\partial x}\right)F(x,t),
\end{equation}
with  $F(x,0)=e^{-\frac{(x-b\chi)^{2}}{2\sigma^{2}(1-b^{2})}}/\sqrt{2\pi\sigma^{2}(1-b^{2})}$ initially. The marginal average for the decoherence function in Eq.~\eqref{eq:maraveOUN} can be solved by means of the approach of eigenfunction expansion~\cite{Riskenbook,PhysRevA105.052443,Moiseyevbook,PhysRevA86.064104,Szegobook}.

\subsubsection{Case of linear dependence}
For the case that the fluctuation of the frequency difference is linearly dependent on the OUN, namely, $k=1$, the solution of the marginal average for the decoherence function in Eq.~\eqref{eq:maraveOUN} can be written as~(see Appendix~\ref{AppB1} for the details of the derivation)
\begin{equation}
	\label{eq:comaveOUNlin}
	\begin{split}
		F(x,t)=&\frac{1}{\sqrt{2\pi\sigma^{2}(1-b^{2}e^{-2\gamma t})}}\exp\bigg\{-\frac{[x-\mu(t)]^{2}}{2\sigma^{2}(1-b^{2}e^{-2\gamma t})}\\
		&+i\frac{cb\chi }{\gamma}\eta(t)-\frac{c^{2}\sigma^{2}}{\gamma^{2}}\left[\gamma t-\eta(t)+\frac{1}{2}b^{2}\eta^{2}(t)\right]\bigg\},
	\end{split}
\end{equation}
where the auxiliary functions $\eta(t)$ and $\mu(t)$ can be expressed as
\begin{equation}
	\label{eq:comaveOUNauxfun}
	\begin{split}
		\eta(t)&=1-e^{-\gamma t},\\
		\mu(t)&=b\chi e^{-\gamma t}+i\frac{c\sigma^{2}}{\gamma}\eta(t)(1+b^{2}e^{-\gamma t}).
	\end{split}
\end{equation}

The decoherence function $F(t)$ can be, by taking the integral of the marginal average $F(x,t)$ over $x$, analytically expressed as
\begin{equation}
	\label{eq:decfunOUNlin}
	\begin{split}
		F(t)&=\exp\left\{i\frac{cb\chi }{\gamma}\eta(t)-\frac{(c\sigma)^{2}}{\gamma^{2}}\left[\gamma t-\eta(t)+\frac{1}{2}b^{2}\eta^{2}(t)\right]\right\}.
	\end{split}
\end{equation}
This expression of the decoherence function in Eq.~\eqref{eq:decfunOUNlin} is consistent with that derived by means of the closed time-convolutionless equation in terms of the time-ordered cumulants~(see Appendix~\ref{sec:AppD.1}).
Therefore, the decoherence rate $\Gamma(t)$ can be expressed as
\begin{equation}
	\label{eq:decratOUNlin}
	\begin{split}
		\Gamma(t)=\frac{(c\sigma)^{2}}{\gamma}(1-e^{-\gamma t})(1+b^{2}e^{-\gamma t}),
	\end{split}
\end{equation}
and the frequency shift $S(t)$ can be  written as 
\begin{equation}
	\label{eq:freshiOUNlin}
	\begin{split}
		S(t)=-cb\chi e^{-\gamma t}.
	\end{split}
\end{equation}

For the stationary case $b=0$ or $\chi=0$, there is no renormalization of the intrinsic energy levels of the system, namely, $S(t)=0$.
For the standard case $b=0$, the decoherence function in Eq.~\eqref{eq:decfunOUNlin} can be simplified as
\begin{equation}
	\label{eq:decfunOUNlinsta}
	\begin{split}
		F(t)&=\exp\left[-\frac{(c\sigma)^{2}}{\gamma^{2}}\left(e^{-\gamma t}+\gamma t-1\right)\right],
	\end{split}
\end{equation}
and the decoherence rate in Eq.~\eqref{eq:decratOUNlin} can be rewritten as
\begin{equation}
	\label{eq:decratOUNlinsta}
	\begin{split}
		\Gamma(t)=\frac{(c\sigma)^{2}}{\gamma}(1-e^{-\gamma t}).
	\end{split}
\end{equation}

\begin{figure}[ht]
	\centering
	\includegraphics[width=3.4in]{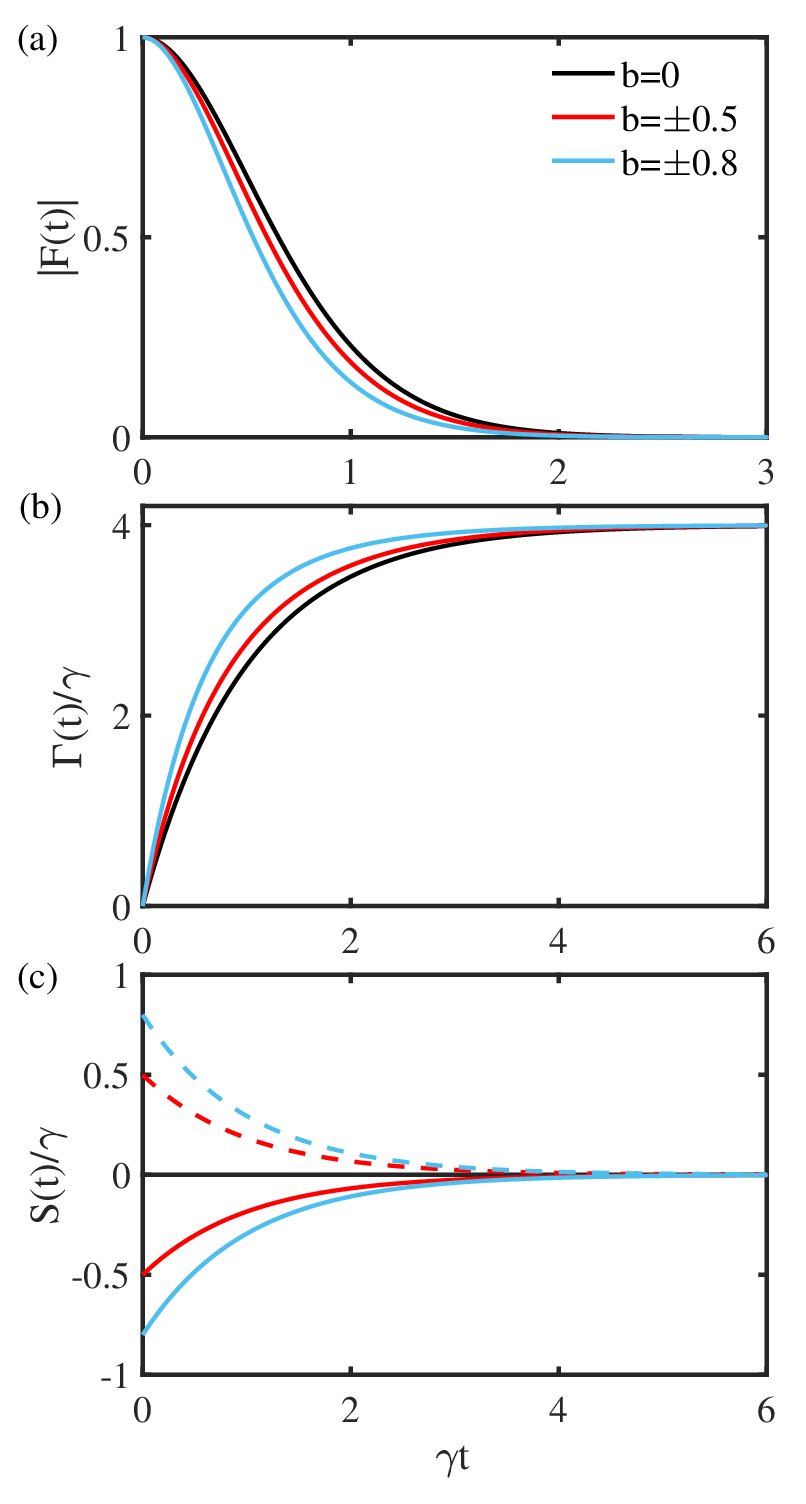}
	\caption{(Color online) Time evolution of  (a) decoherence function $|F(t)|$, (b) decoherence rate $\Gamma(t)$ and (c) frequency shift $S(t)$ (the solid lines for $b > 0$ and the dashed lines are for $b< 0$) in the case of linear dependence of OUN for different values of  nonstationary parameter $b$.
		The parameters are set as $c=1$, $\sigma=2\gamma$ and $\chi=\gamma$.}
	\label{fig:OUNlin}
\end{figure}

As depicted in Fig.~\ref{fig:OUNlin}~(a), $|F(t)|$ decreases monotonically as a function of time arising from the Gaussian feature of the environmental noise for both the stationary and nonstationary cases.
For a given time $t$, $|F(t)|$ decreases with the increase of the nonstationary parameter $|b|$.
This indicates that the nonstationary statistical property of the OUN can enhance the dynamical decoherence in the case of the fluctuation term under the linear dependence of the environmental noise.

The decoherence rate $\Gamma(t)$ in Eq.~\eqref{eq:decratOUNlin} increases monotonically as a function of evolution time $t$ with $\Gamma(0)=0$ since 
\begin{equation}
	\label{eq:decratOUNlindt}
	\begin{split}
		\frac{d}{dt}\Gamma(t)=(c\sigma)^{2}e^{-\gamma t}\left[(1-b^{2})+2b^{2}e^{-2\gamma t}\right]>0,
	\end{split}
\end{equation}
for any time $t>0$. This behavior can be clearly seen in Fig.~\ref{fig:OUNlin}~(b).
In the long time limit, the decoherence rate approaches the steady value $\Gamma_{s}=(c\sigma)^{2}/\gamma$ which is independent of the nonstationary parameter $|b|$.
The decoherence rate $\Gamma(t)$ for the nonstationary case $b\neq0$ is always larger than that for the standard case $b=0$ at any time $t>0$. With the increase of $|b|$, the decoherence rate in Eq.~\eqref{eq:decratOUNlin} increases for a given evolution time $t>0$ since
\begin{equation}
	\label{eq:decratOUNlindb}
	\begin{split}
		\frac{d}{db}\Gamma(t)=2b\frac{(c\sigma)^{2}}{\gamma}e^{-\gamma t}(1-e^{-\gamma t})>0,
	\end{split}
\end{equation}
for $0<b<1$. 
It is just the reason why the nonstationary statistics of the OUN can enhance the dynamical decoherence of the quantum system.

For the nonstationary case $b\neq0$ and $\chi\neq0$, the frequency shift $S(t)$ is a monotonic function of time $t$ with $S(0)=-cb\chi$ and it vanishes in the long time limit $t\rightarrow\infty$.
For a given $\chi\neq0$, $S(t)$ is symmetric with respect to the $t$-axis for positive and negative values of $cb$.
For a given time $t$, $|S(t)|$ increases with the increase of the value of $|b|$ which suggests that the nonstationary statistics of the OUN can heighten the frequency renormalization of the system.
This fact is illustrated in Fig.~\ref{fig:OUNlin}~(c).

\subsubsection{Case of quadratic dependence}
For the case that the fluctuation of the frequency difference quadratically depends on the OUN, namely, $k=2$, the marginal average for the decoherence function in Eq.~\eqref{eq:maraveOUN} can be solved as~(see Appendix~\ref{AppB2} for the details of the derivation)
\begin{equation}
	\label{eq:comaveOUNqua}
	\begin{split}	
		F(x,t)=&\sqrt{\frac{\epsilon}{\pi\sigma^{2}A(t)}}\exp\bigg\{\frac{\ell_{-}(\epsilon-1)e^{-2\epsilon\gamma t}-\ell_{+}(\epsilon+1)}{4\sigma^{2}A(t)}x^{2}\\
		&+\frac{2b\chi\epsilon e^{-\epsilon\gamma t}}{\sigma^{2}A(t)}x-\frac{(b\chi)^{2}[(\epsilon+1)e^{-2\epsilon\gamma t}+\epsilon-1]}{2\sigma^{2}A(t)}\\
		&-\frac{\gamma}{2}(\epsilon-1) t\bigg\},
	\end{split}	
\end{equation}
where $\epsilon=\sqrt{1-4ic\sigma^{2}/\gamma}$, $\ell_{\pm}=(1-b^{2})\epsilon\pm(1+b^{2})$ and $A(t)=\ell_{+}+\ell_{-}e^{-2\epsilon\gamma t}$.
Therefore, by integrating the marginal average $F(x,t)$ over $x$, we obtain the analytical expression of the decoherence function $F(t)$ as
\begin{equation}
	\label{eq:decfunOUNqua}
	\begin{split}	
		F(t)=&\sqrt{\frac{4\epsilon}{\ell_{+}(\epsilon+1)-\ell_{-}(\epsilon-1)e^{-2\epsilon\gamma t}}}\exp\bigg\{-\frac{\gamma}{2}(\epsilon-1)t\\
		&-\frac{(b\chi)^{2}(\epsilon^{2}-1)(1-e^{-2\epsilon\gamma t})}{2\sigma^{2}[\ell_{+}(\epsilon+1)-\ell_{-}(\epsilon-1)e^{-2\epsilon\gamma t}]}\bigg\}.
	\end{split}	
\end{equation}
Consequently, the decoherence rate $\Gamma(t)$ can be written as
\begin{equation}
	\label{eq:decratOUNnonlin}
	\begin{split}
		\Gamma(t)&=\frac{\gamma\left[C(t)e^{-2\alpha\gamma t}+\alpha d_{+}\right]}{d_{+}-2B(t)e^{-2\alpha\gamma t}+g_{+}e^{-4\alpha\gamma t}}+\frac{2\alpha\beta\gamma (b\chi)^{2}}{\sigma^{2}e^{2\alpha\gamma t}}\\
		&\quad\times\frac{I(t)-2\beta Je^{-2\alpha\gamma t}+K(t)e^{-4\alpha\gamma t}}{\left[d_{+}-2B(t)e^{-2\alpha\gamma t}+g_{+}e^{-4\alpha\gamma t}\right]^{2}}-\frac{\gamma}{2}(\alpha+1),
	\end{split}
\end{equation}
and the frequency shift $S(t)$ can be expressed as 
\begin{equation}
	\label{eq:freshiOUNnonlin}
	\begin{split}
		S(t)&=\frac{\gamma\left[D(t)e^{-2\alpha\gamma t}-\beta d_{+}\right]}{d_{+}-2B(t)e^{-2\alpha\gamma t}+g_{+}e^{-4\alpha\gamma t}}+\frac{2\alpha\beta\gamma (b\chi)^{2}}{\sigma^{2}e^{2\alpha\gamma t}}\\
		&\quad\times\frac{L(t)+2Me^{-2\alpha\gamma t}+N(t)e^{-4\alpha\gamma t}}{\left[d_{+}-2B(t)e^{-2\alpha\gamma t}+g_{+}e^{-4\alpha\gamma t}\right]^{2}}+\frac{\gamma}{2}\beta.
	\end{split}
\end{equation}
Here $B(t)=\beta h_{+}\sin(2\beta\gamma t)+j_{-}\cos(2\beta\gamma t)$, the other auxiliary parameters in Eq.~\eqref{eq:decratOUNnonlin} can be expressed as
\begin{equation}
	\label{eq:auxfunGam}
	\begin{split}
		C(t)=&\beta(j_{-}-\alpha h_{+})\sin(2\beta\gamma t)-(\alpha j_{-}+\beta^{2} h_{+})\cos(2\beta\gamma t),\\
		I(t)=&(d_{-}+4\alpha\beta^{2}u)\sin(2\beta\gamma t)+2\beta(u-\alpha d_{-})\cos(2\beta\gamma t),\\
		J=&h_{-}+2\alpha j_{+},\\
		K(t)=&(4\alpha\beta^{2}v- g_{-})\sin(2\beta\gamma t)-2\beta(\alpha g_{-}+v)\cos(2\beta\gamma t),
	\end{split}
\end{equation}
and the other auxiliary parameters in Eq.~\eqref{eq:freshiOUNnonlin} can be written as
\begin{equation}
	\label{eq:auxfunS}
	\begin{split}
		D(t)=&(\alpha j_{-}+\beta^{2} h_{+})\sin(2\beta\gamma t)+\beta(j_{-}-\alpha h_{+})\cos(2\beta\gamma t),\\
		L(t)=&2\beta(u-\alpha d_{-})\sin(2\beta\gamma t)-(d_{-}+4\alpha\beta^{2}u)\cos(2\beta\gamma t),\\
		M=&2\alpha\beta^{2}h_{-}-j_{+},\\
		N(t)=&2\beta(\alpha g_{-}+v)\sin(2\beta\gamma t)+(4\alpha\beta^{2}v- g_{-})\cos(2\beta\gamma t),
	\end{split}
\end{equation}
where $u=(\alpha+1)[\alpha(1-b^{2})+1]$, $v=(\alpha-1)[\alpha(1-b^{2})-1]$ and
\begin{equation}
	\label{eq:auxpar}
	\begin{split}
		d_{\pm}&=(\alpha+1)^{2}\pm\beta^{2}\left[\alpha(1-b^{2})+1\right]^{2},\\
		g_{\pm}&=(\alpha-1)^{2}\pm\beta^{2}\left[\alpha(1-b^{2})-1\right]^{2},\\
		h_{\pm}&=(\alpha+1)[\alpha(1-b^{2})-1]\pm(\alpha-1)[\alpha(1-b^{2})+1],\\
		j_{\pm}&=\beta^{2}[\alpha^{2}(1-b^{2})^{2}-1]\pm(\alpha^{2}-1),
	\end{split}
\end{equation}
with 
\begin{equation}
	\label{eq:alpbet}
	\begin{split}
		\alpha&=\sqrt{\frac{\sqrt{1+16c^{2}\sigma^{4}/\gamma^{2}}+1}{2}},\\
		\beta&=\sqrt{\frac{\sqrt{1+16c^{2}\sigma^{4}/\gamma^{2}}-1}{2}}.
	\end{split}
\end{equation}

For the standard stationary case $b=0$, the decoherence function in Eq.~\eqref{eq:decfunOUNqua} can be further reduced to 
\begin{equation}
	\label{eq:comavedecfunOUTCLSecC}
	\begin{split}
		F(t)=&\sqrt{\frac{4\epsilon}{(\epsilon+1)^{2}-(\epsilon-1)^{2}e^{-2\epsilon\gamma t}}}\exp\left[-\frac{\gamma}{2}(\epsilon-1) t\right].
	\end{split}
\end{equation}
The decoherence rate can be rewritten as
\begin{equation}
	\label{eq:decratOUNnonlinsta}
	\begin{split}
		\Gamma(t)=&\frac{\gamma\left[\beta^{2}U(t)e^{-2\alpha\gamma t}+\alpha^{3}(\alpha+1)^{2}\right]}{[\alpha(\alpha+1)]^{2}-2\beta^{2}E(t)e^{-2\alpha\gamma t}+[\alpha(\alpha-1)]^{2}e^{-4\alpha\gamma t}}\\
		&-\frac{\gamma}{2}(\alpha+1),
	\end{split}
\end{equation}
and the frequency renormalization still exists
\begin{equation}
	\label{eq:freshiOUNnonlinsta}
	\begin{split}
		S(t)=&\frac{\gamma\beta\left[\beta V(t)e^{-2\alpha\gamma t}-\alpha^{2}(\alpha+1)^{2}\right]}{[\alpha(\alpha+1)]^{2}-2\beta^{2}E(t)e^{-2\alpha\gamma t}+[\alpha(\alpha-1)]^{2}e^{-4\alpha\gamma t}}\\
		&+\frac{\gamma}{2}\beta,
	\end{split}
\end{equation}
where the auxiliary parameters are given by
\begin{equation}
	\label{eq:decratOUNnonlinstapar}
	\begin{split}
		E(t)&=2\beta\sin(2\beta\gamma t)+(\beta^{2}-1)\cos(2\beta\gamma t),\\
		U(t)&=\theta\sin(2\beta\gamma t)-\vartheta\cos(2\beta\gamma t),\\
		V(t)&=\vartheta\sin(2\beta\gamma t)+\theta\cos(2\beta\gamma t),
	\end{split}
\end{equation}
with $\theta=\beta\left[(\alpha-1)^{2}-3\right]$ and $\vartheta=\alpha^{2}(\alpha+2)-2(\alpha+1)$.

\begin{figure}[ht]
	\centering
	\includegraphics[width=3.4in]{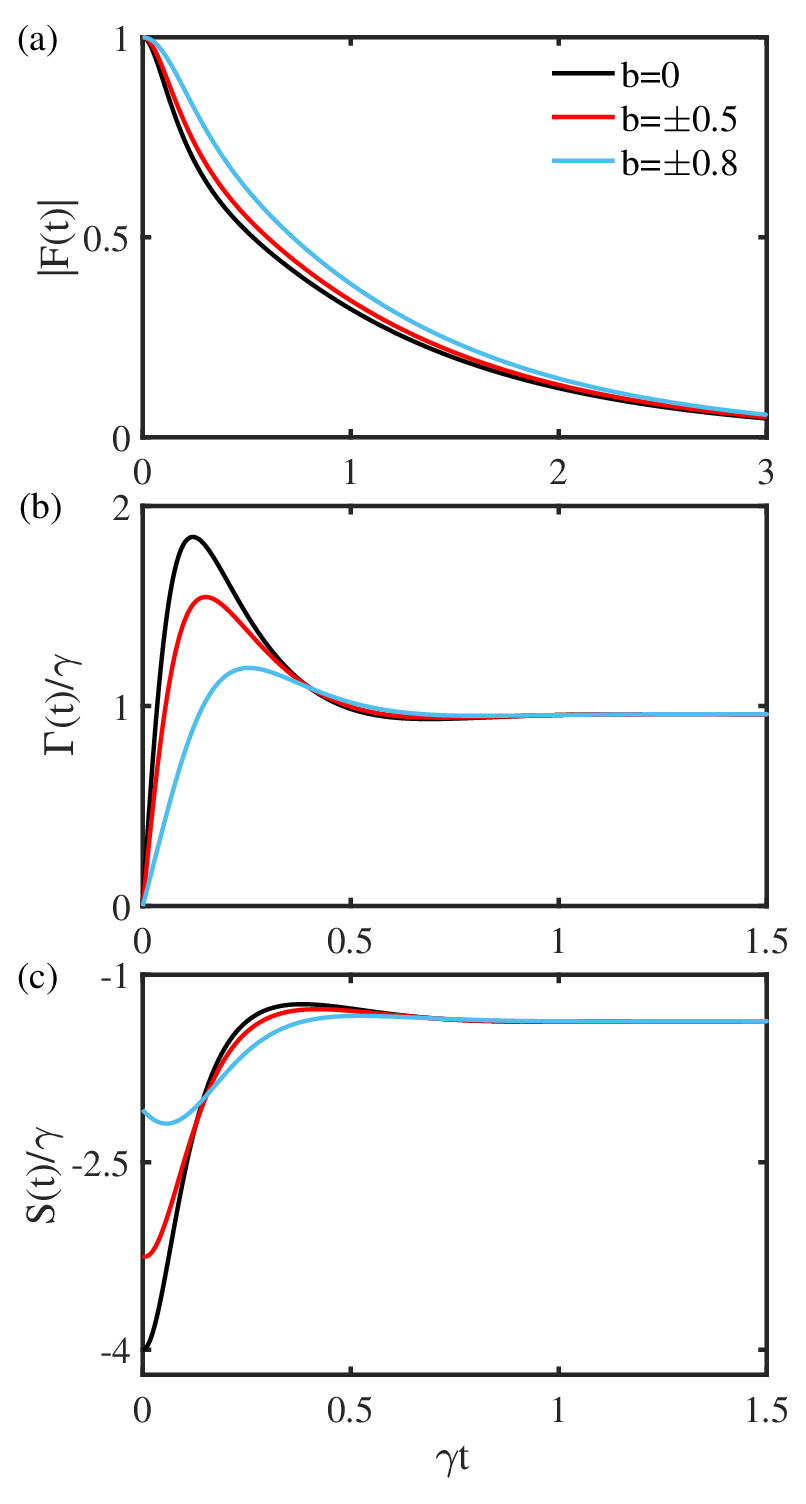}
	\caption{(Color online) Time evolution of  (a) decoherence function $|F(t)|$, (b) decoherence rate $\Gamma(t)$ and (c) frequency shift $S(t)$ in the case of quadratic dependence of OUN for different values of  nonstationary parameter $b$.
		The parameters are chosen as $|c|=1$, $\sigma=2\gamma$ and $|\chi|=\gamma$.}
	\label{fig:OUNnonlin}
\end{figure}

As  illustrated in Fig.~\ref{fig:OUNnonlin}~(a), $|F(t)|$ is a monotone decreasing function of time for both the stationary and nonstationary cases due to the Gaussian feature of the OUN.
For a given time $t$, $|F(t)|$ increases with the increase of the nonstationary parameter $|b|$.
This means that the environmental nonstationary statistical property can reduce the dynamical decoherence in the case of the fluctuation term under the quadratic influence of the OUN.

Figure~\ref{fig:OUNnonlin}~(b) shows the decoherence rate $\Gamma(t)$ as a function of time $t$ for different valves of $|b|$.
For both the stationary and nonstationary cases, $\Gamma(t)$ increases nonmonotonically from the initial value $\Gamma(0)=0$ as time evolves and it approaches the steady value $\Gamma_{s}=\gamma(\alpha-1)/2$ independent of $|b|$ in the long time limit.
At a given short time $t$, $\Gamma(t)$ decreases with the increase of the value of $|b|$.
This is the main reason why the dynamical decoherence gets reduced by the nonstationary statistical property of the OUN in the case of the quadratic dependence of the environmental noise.

As depicted in Fig.~\ref{fig:OUNnonlin}~(c), the frequency shift $S(t)$ always emerges with $S(0)=-\alpha\beta\gamma[(1-b^{2})+(b\chi)^{2}/\sigma^{2}]/2$ and it reaches the same constant value $S_{s}=-\gamma\beta/2$ for different valves of $b$ as time approaches infinity for both the stationary and nonstationary cases.
In short times, $|S(t)|$ decreases with the increase of the value of $|b|$. This reflects the nonstationary statistical property of the OUN can suppress the frequency renormalization of the system in the case of the fluctuation term under the quadratic influence of environmental noise.

\subsubsection{Comparison between the cases in the presence of linear and nonlinear OUN}

\begin{figure}[ht]
	\centering
	\includegraphics[width=3.4in]{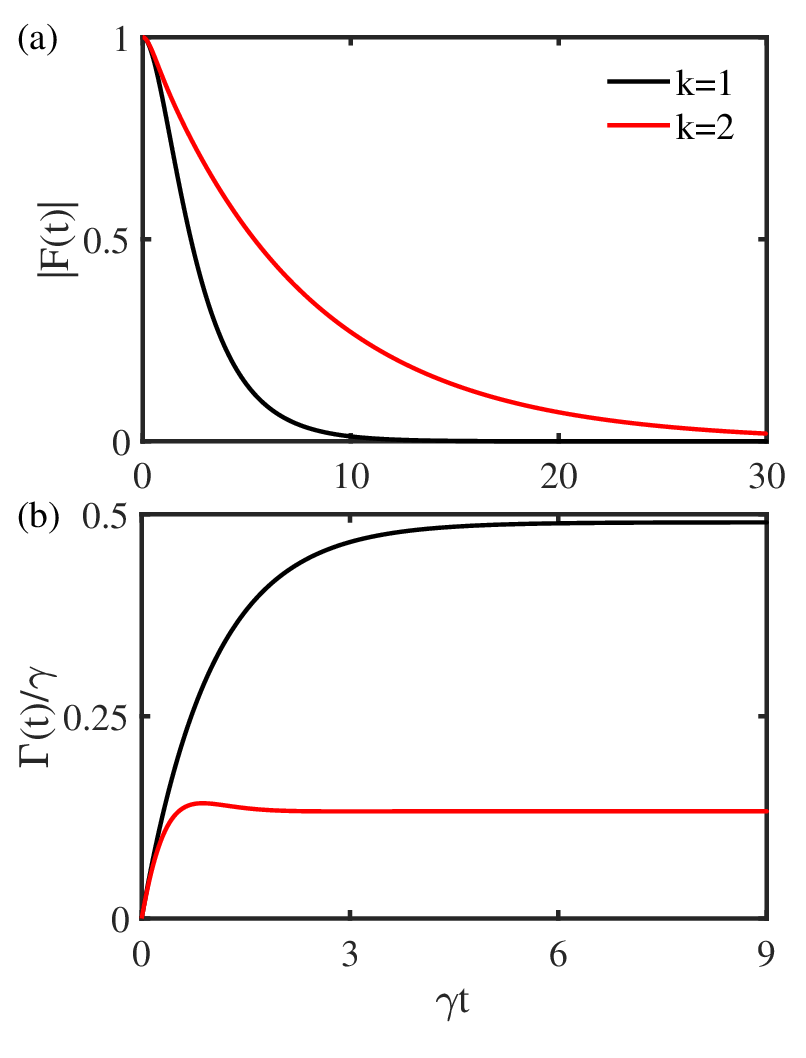}
	\caption{(Color online) Time evolution of (a) decoherence function $|F(t)|$ and (b) decoherence rate $\Gamma(t)$ for the cases of linear ($k=1$) and quadratic ($k=2$) dependence of OUN in weak coupling regime.
		The parameters are chosen as $|c|=1$ and $\sigma=0.7\gamma$.}
	\label{fig:OUNFtGammat1}
\end{figure}

\begin{figure}[ht]
	\centering
	\includegraphics[width=3.4in]{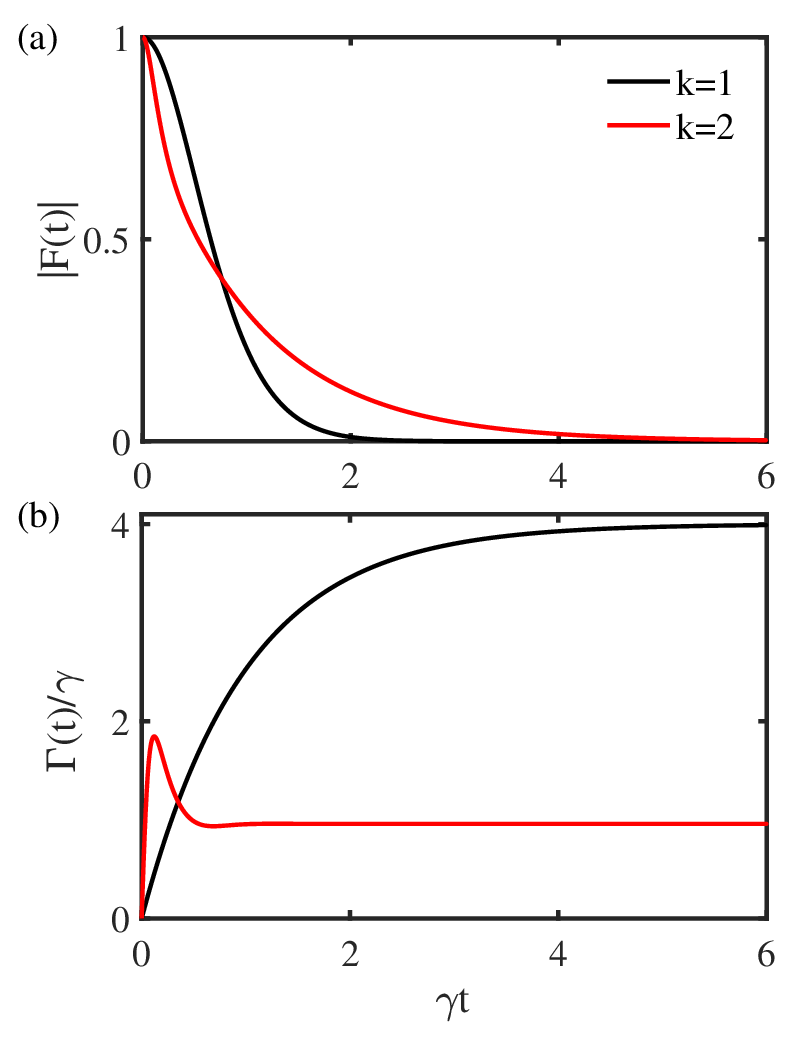}
	\caption{(Color online) Time evolution of (a) decoherence function $|F(t)|$ and (b) decoherence rate $\Gamma(t)$ in the cases of linear and quadratic dependence of standard OUN in the strong coupling regime.
		The parameters are chosen as $|c|=1$ and $\sigma=2\gamma$.}
	\label{fig:OUNFtGammat2}
\end{figure}

Comparing the results of the linear and quadratic fluctuations in the frequency difference governed by the OUN, we can find that the decoherence function decays exponentially which is independent of the environmental nonstationary statistics for both the linear and nonlinear cases in the long time limit due to the Gaussian feature of the environmental noise. 
It is shown that in the case of linear environmental fluctuations, the frequency renormalization of the system only emerges when the OUN exhibits nonstationary statistics and vanishes in long time limit whereas it always emerges even when the OUN exhibits stationary statistical properties and does not vanish as time goes to infinity in the case of quadratic fluctuations.
For short times, the environmental nonstationary statistical property plays the negative and positive role in suppressing dynamical decoherence in the cases of linear and nonlinear fluctuations, respectively.

To further make a comparison on the dynamical decoherence between the cases of linear and quadratic  environmental fluctuations, we plot the time evolution of the decoherence function $|F(t)|$ and the decoherence rate $\Gamma(t)$ for the standard stationary OUN in the weak and strong coupling regimes depicted in Figs.~\ref{fig:OUNFtGammat1} and~\ref{fig:OUNFtGammat2}, respectively.
In the weak coupling regime, the decoherence function $|F(t)|$ for the quadratic  environmental fluctuations is always larger than that for the linear case as shown in Fig.~\ref{fig:OUNFtGammat1} (a) and the decoherence rate $\Gamma(t)$ for the quadratic case is always smaller than the case in the presence of linear environmental fluctuations as displayed in Fig.~\ref{fig:OUNFtGammat1} (b).
This indicates that the nonlinear dependence of the environmental fluctuations can suppress the dynamical decoherence in the weak coupling regime.
Figure~\ref{fig:OUNFtGammat2} (a) shows the comparison of the decoherence function $|F(t)|$ between the linear and quadratic cases in the strong coupling regime.
In a short time, the decoherence function for the nonlinear case is smaller than that of linear environmental fluctuations.
As the time evolves beyond a critical value $t_{\mathrm{cr}}$, it begins to be larger than that for the linear case.
This is mainly due to that the nonlinear environmental fluctuations make the decoherence rate $\Gamma(t)$ increase rapidly in a short time and reach a smaller steady value in long time limit than that for the linear case as depicted in Fig.~\ref{fig:OUNFtGammat2} (b).
This implies that compared to the case in the presence of linear environmental fluctuations, the nonlinear influence of the OUN can enhance the dynamical decoherence in short time and suppress the quantum decoherence in a long time in the strong coupling regime.

\subsection{Decoherence in the presence of environmental noise governed by nonstationary RTN}
The standard RTN as an important stationary non-Gaussian process has been widely used to study the environmental influences on open quantum systems, such as, the single molecule fluorescence, disentanglement, decoherence and frequency modulation processes~\cite{NewJPhys.8.1,PhysRevA73.022332,PhysRevB75.054515,JChemPhys.144.024113,PhysRevA100.052104,PhysRevA107.L030601,ApplPhysLett.122.244001}.

We now consider the case that the environmental noise $x(t)$ is governed by the RTN, which exhibits nonstationary statistics.
The statistical characteristics of the nonstationary RTN can be extracted from those of the standard one with stationary statistics~(see Appendix~\ref{sec:AppA}).
For the nonstationary RTN, the transition probability satisfies the master equation
\begin{equation}
	\label{eq:traproRTN}
	\begin{split}
		\frac{\partial}{\partial t}P(\nu,t|x',t')&=-\lambda P(\nu,t|x',t')+\lambda P(-\nu,t|x',t'),\\
		\frac{\partial}{\partial t}P(-\nu,t|x',t')&=-\lambda P(-\nu,t|x',t')+\lambda P(\nu,t|x',t'),
	\end{split}
\end{equation}
where the switching rate $\lambda$ quantifies the spectral width of the environmental coupling, the transition amplitude $\nu$ characterizes the coupling strength between the system and environment and the initial condition yields $P(x,t|x',t')=\delta_{x,x'}$ for $x=\pm\nu$.
For the environment in a nonstationary initial state with RTN statistics, the nonstationary statistical property of the environmental noise is described by the initial single-point probability
\begin{equation}
	\label{eq:inisinpoiRTN}
	P(x,0)=\frac{1}{2}(1+a)\delta_{x,\nu}+\frac{1}{2}(1-a)\delta_{x,-\nu},
\end{equation}
with the initial nonstationary parameter $|a|\leq1$.
For the case $a=0$, the RTN returns to the standard one with stationary statistical properties.

Under the influence of environmental noise subject to the nonstationary RTN, the master equation of the marginal average for the decoherence function in Eq.~\eqref{eq:masequparave} can be further expressed as 
\begin{equation}
	\label{eq:maraveRTN}
	\begin{split}
		\frac{\partial}{\partial t}F(\nu,t)&=ic\nu^{k} F(\nu,t)-\lambda F(\nu,t)+\lambda F(-\nu,t),\\
		\frac{\partial}{\partial t}F(-\nu,t)&=ic(-\nu)^{k} F(-\nu,t)-\lambda F(-\nu,t)+\lambda F(\nu,t),
	\end{split}
\end{equation}
with the initial conditions $F(\pm\nu,0)=(1\pm a)/2$.
This is a special case where the environmental noise $x(t)$ takes two discrete states $\pm\nu$.
The marginal average for the decoherence function in Eq.~\eqref{eq:maraveRTN} can be solved by means of the Laplace transform technique.

\subsubsection{Case of linear dependence}
For the case that the fluctuation term of the frequency difference linearly depends on the RTN, namely, $k=1$, the marginal average for the decoherence function in Eq.~\eqref{eq:maraveRTN} can be solved as~(see Appendix~\ref{AppC1} for the details of the derivation)
\begin{equation}
	\label{eq:maraveRTNlin}
	\begin{split}	
		F(\nu,t)&=\frac{1}{2}e^{-\lambda t}\bigg[(1+a)\cosh(\kappa t)+(1-a)\frac{\lambda}{\kappa}\sinh(\kappa t)\\
		&\quad+i(1+a)\frac{c\nu}{\kappa}\sinh(\kappa t)\bigg],\\
		F(-\nu,t)&=\frac{1}{2}e^{-\lambda t}\bigg[(1-a)\cosh(\kappa t)+(1+a)\frac{\lambda}{\kappa}\sinh(\kappa t)\\
		&\quad-i(1-a)\frac{c\nu}{\kappa}\sinh(\kappa t)\bigg],
	\end{split}
\end{equation}
with the parameter $\kappa=\sqrt{\lambda^{2}-(c\nu)^{2}}$.

Therefore, the decoherence function $F(t)$ can be expressed as the sum of the marginal averages $F(\pm\nu,t)$ as
\begin{equation}
	\label{eq:decfunRTNlin}
	F(t)=e^{-\lambda t}\left[\cosh(\kappa t)+\frac{\lambda}{\kappa}\sinh(\kappa t)+i\frac{ac\nu}{\kappa}\sinh(\kappa t)\right].
\end{equation}
The analytical expression of the decoherence function in Eq.~\eqref{eq:decfunRTNlin} consists with that derived by means of the closed dynamical equation in the time-convolution form with an inhomogeneous term~(see Appendix~\ref{sec:AppD.2}).
Correspondingly, the decoherence rate $\Gamma(t)$ can be expressed as 
\begin{equation}
	\label{eq:decratRTNlin}
	\Gamma(t)=\frac{(c\nu)^{2}\left[\frac{\kappa}{2}(1-a^{2})\sinh(2\kappa t)+\lambda(1+a^{2})\sinh^{2}(\kappa t)\right]}{\left[\kappa\cosh(\kappa t)+\lambda\sinh(\kappa t)\right]^{2}+[ac\nu\sinh(\kappa t)]^{2}},
\end{equation}
and the frequency shift $S(t)$ is expressed as
\begin{equation}
	\label{eq:freshiRTNlin}
	S(t)=-\frac{ac\nu\kappa^{2}}{\left[\kappa\cosh(\kappa t)+\lambda\sinh(\kappa t)\right]^{2}+\left[ac\nu\sinh(\kappa t)\right]^{2}}.
\end{equation}

For the standard case $a=0$, there is no frequency renormalization induced by the environmental noise, namely, $S(t)=0$. 
The decoherence function in Eq.~\eqref{eq:decfunRTNlin} can be further simplified as
\begin{equation}
	\label{eq:decfunRTNlinsta}
	F(t)=e^{-\lambda t}\left[\cosh(\kappa t)+\frac{\lambda}{\kappa}\sinh(\kappa t)\right],
\end{equation}
and the decoherence rate in Eq.~\eqref{eq:decratRTNlin} can be reduced to
\begin{equation}
	\label{eq:decratRTNlinsta}
	\Gamma(t)=\frac{(c\nu)^{2}\sinh(\kappa t)}{\kappa\cosh(\kappa t)+\lambda\sinh(\kappa t)}.
\end{equation}

\begin{figure}[ht]
	\centering
	\includegraphics[width=3.4in]{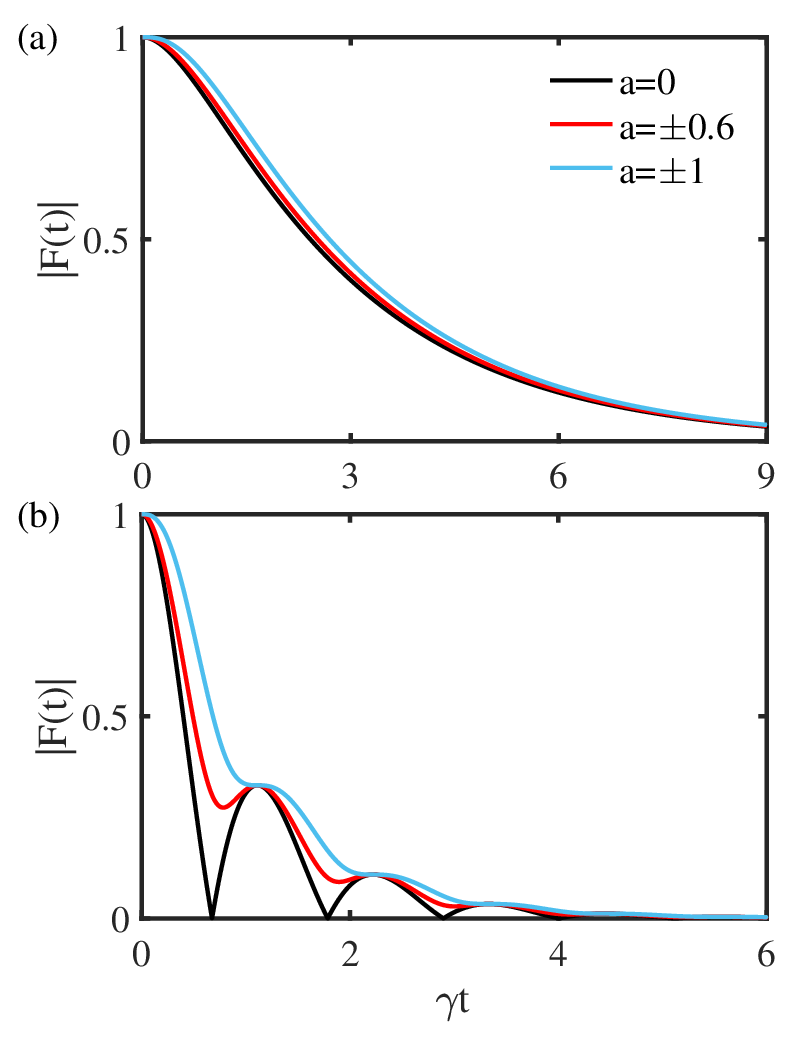}
	\caption{(Color online) Time evolution of the decoherence function $|F(t)|$ in the case of linear dependence of RTN for different values of  nonstationary parameter $a$ in (a) weak coupling regime $|c|/r=0.8$ and (b) strong coupling regime $|c|/r=3$.}
	\label{fig:RTNlin}
\end{figure}

There are two dynamics regimes that can be identified in terms of the ratio $r=\lambda/\nu$: the weak coupling with $|c|<r$, where $\kappa$ is real and the strong coupling $|c|>r$, where $\kappa$ is purely imaginary, respectively.
As shown in Fig.~\eqref{fig:RTNlin} (a), $|F(t)|$ decreases monotonically with time in the weak coupling regime. In the strong coupling regime, $|F(t)|$ is a nonmonotonic decreasing function of time for $|a|<1$ whereas it decreases monotonically as a function of time for $|a|=1$ as depicted in Fig.~\eqref{fig:RTNlin} (b). 

\begin{figure}[ht]
	\centering
	\includegraphics[width=3.4in]{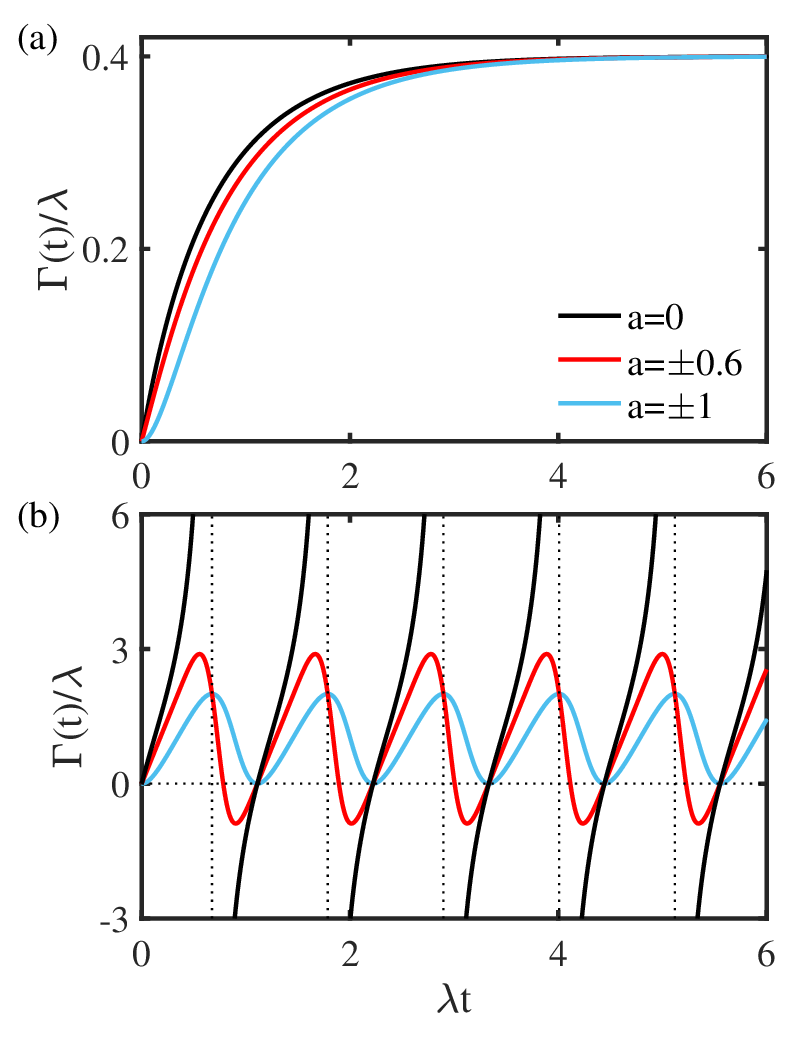}
	\caption{(Color online) Time evolution of  the decoherence rate $\Gamma(t)$ in the case of linear dependence of RTN for different values of  nonstationary parameter $a$ in (a) weak coupling regime $|c|/r=0.8$ and (b) strong coupling regime $|c|/r=3$.}
	\label{fig:RTNlinGammat}
\end{figure}

In the weak coupling regime, the decoherence rate $\Gamma(t)$ in Eq.~\eqref{eq:decratRTNlin} is a monotonic increasing function of evolution time $t$ with $\Gamma(0)=0$ since for any time $t>0$
\begin{equation}
	\label{eq:decratRTNlindt}
	\begin{split}
		\frac{d}{dt}\Gamma(t)=\frac{(c\nu\kappa)^{2}G(t)}{\sinh^{2}(\kappa t)\left\{\left[\kappa\coth(\kappa t)+\lambda\right]^{2}+(ac\nu)^{2}\right\}^2}>0,
	\end{split}
\end{equation}
with the time-dependent auxiliary parameter
\begin{equation}
	\label{eq:decratRTNlindtJt}
	\begin{split}
		G(t)&=\kappa^{2}(1-a^{2})\coth^{2}(\kappa t)+2\lambda\kappa(1+a^{2})\coth(\kappa t)\\
		&\quad+\lambda^{2}(1+a^{2})^{2}+\kappa^{2}a^{2}(1-a^{2}).
	\end{split}
\end{equation}
The decoherence rate approaches the steady value $\Gamma_{s}=(c\nu)^{2}/(\lambda+\kappa)$ which is independent of $a$ in the long time limit.
For a given evolution time $t>0$, the decoherence rate $\Gamma(t)$ for the nonstationary case $a\neq0$ is always smaller than that for the standard case $a=0$.  
The decoherence rate in Eq.~\eqref{eq:decratRTNlin} decreases with the increase of the initial nonstationary parameter $|a|$ since
\begin{equation}
	\label{eq:decratRTNlinda}
	\begin{split}
		\frac{d}{da}\Gamma(t)=-\frac{2a(c\nu\kappa)^{2}[\kappa\cosh(\kappa t)+\lambda][\coth^{2}(\kappa t)-1]}{\left\{\left[\kappa\coth(\kappa t)+\lambda\right]^{2}+(ac\nu)^{2}\right\}^2}<0,
	\end{split}
\end{equation}
for $0<a\leq1$. This implies that the nonstationary statistical property of the RTN can reduce the dynamical decoherence in the weak coupling regime in the case of the fluctuation term linearly depending on the environmental noise.
This fact can be obviously seen in Fig.~\ref{fig:RTNlinGammat}~(a).

In the strong coupling regime, the decoherence rate $\Gamma(t)$ is a periodic function of time with the period $T=\pi/\kappa'$, where $\kappa'=-i\kappa$.
For the stationary case $a=0$, the decoherence rate $\Gamma(t)$ has discrete zeros at
$t_{n}^{0}=(n-1)\pi/\kappa'$, where the integer $n=1,2,\cdots$.
However, it is discontinuous in time and diverges at the time points
\begin{equation}
	\label{eq:decratRTNdiszer}
	\begin{split}
		t_{n}^{d}=\frac{1}{\kappa'}\left(n\pi-\arctan\frac{\kappa'}{\lambda}\right).
	\end{split}
\end{equation}
The decoherence rate $\Gamma(t)$ is positive on time intervals $(t_{n}^{0},t_{n}^{d})$ and it takes negative values for $t\in(t_{n}^{d},t_{n+1}^{0})$.
For the nonstationary case $a\neq0$, the decoherence rate $\Gamma(t)$ is a continuous function of time $t$. 
For the case $|a|<1$, the decoherence rate takes positive and negative values respectively on time intervals $(t_{n}^{0},t'_{n})$ and $(t'_{n},t_{n+1}^{0})$, where 
\begin{equation}
	\label{eq:decratRTNdiszersa}
	\begin{split}
		t'_{n}=\frac{1}{\kappa'}\left[n\pi-\arctan\frac{\kappa'(1-a^{2})}{\lambda(1+a^{2})}\right].
	\end{split}
\end{equation}
While for the case $|a|=1$, it is always non-negative with the period $T=\pi/(2\kappa')$ and the discrete zeros at $t_{n}^{0}$
since $t'_{n}=t_{n+1}^{0}$.
With the increase of the value of $|a|$, the interval length $\tau_{p}=t'_{n}-t_{n}^{0}$ for $\Gamma(t)>0$ increases while the interval length $\tau_{n}=t_{n+1}^{0}-t'_{n}$ for $\Gamma(t)<0$ decreases.
This reflects that the nonstationary statistical property of the RTN can suppress the coherence revivals in dynamical decoherence in the strong coupling regime in the case of the fluctuation term under the linear dependence of the environmental noise.
This behavior is illustrated in Fig.~\ref{fig:RTNlinGammat}~(b).

\begin{figure}[ht]
	\centering
	\includegraphics[width=3.4in]{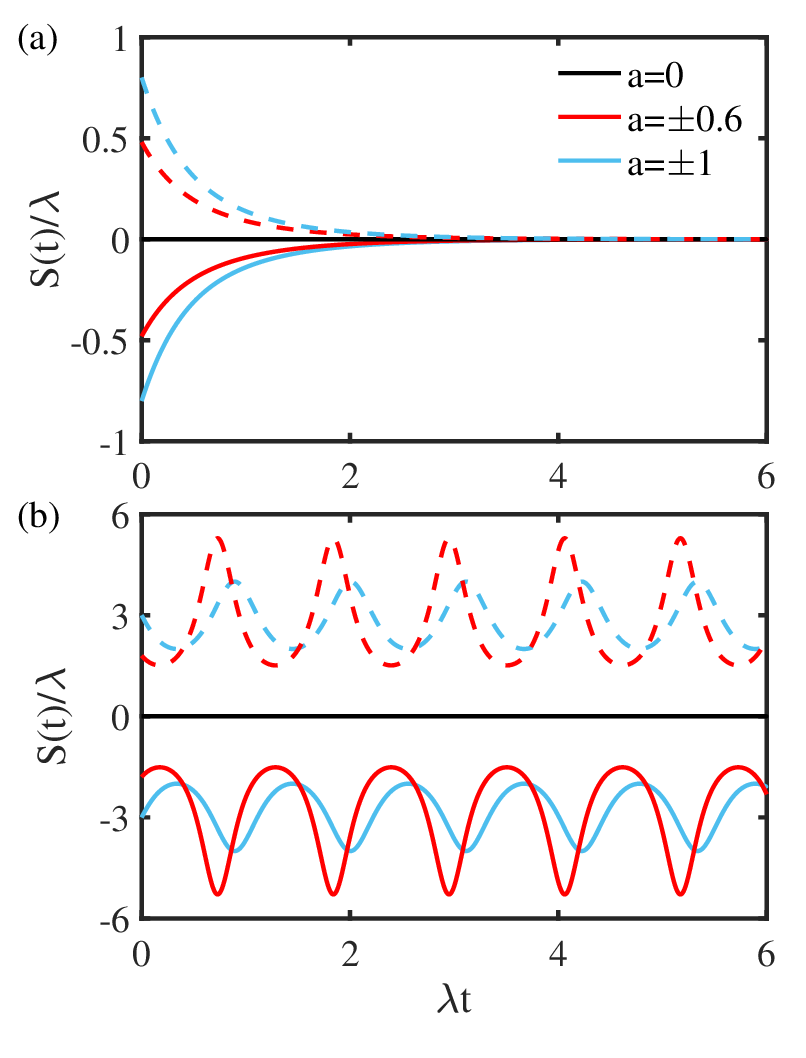}
	\caption{(Color online) Time evolution of  the frequency shift $S(t)$ in the case of linear dependence of RTN for different values of nonstationary parameter $a$ in (a) weak coupling regime $|c|/r=0.8$ and (b) strong coupling regime $|c|/r=3$ (the solid lines are for $ac > 0$ and the dashed lines are for $ac< 0$).}
	\label{fig:RTNlinSt}
\end{figure}

For the nonstationary case $a\neq0$, $|S(t)|$ is monotonically decreasing as a function of time with $|S(0)|=|ac|\nu$ since $(d/dt)|S(t)|>0$ and it vanishes in the long time limit $t\rightarrow\infty$ in the weak coupling regime whereas in the strong coupling regime, it is an oscillating periodic function of time $t$ with the period $T=\pi/\kappa'$.
As the value of $|a|$ increases, $|S(t)|$ increases for short times in the weak coupling regime while its amplitude decreases in the strong coupling regime as depicted in Fig.~\ref{fig:RTNlinSt} (a) and (b).
This means that the nonstationary statistical property of the RTN can heighten the  frequency renormalization of the system in the weak coupling regime whereas it can reduce the frequency shift in the strong coupling regime.

\subsubsection{Case of quadratic dependence}
For the case that the fluctuation of the frequency difference is quadratically dependent on the RTN, namely, $k=2$, the solution of the marginal averages $F(\pm\nu,t)$ in Eq.~\eqref{eq:maraveRTN} can be written as~(see Appendix~\ref{AppC2} for the details of the derivation)
\begin{equation}
	\label{eq:maraveRTNqua}
	\begin{split}
		F(\pm\nu,t)&=\frac{1}{2}\left(1\pm ae^{-2\lambda t}\right)e^{ic\nu^{2}t}.
	\end{split}
\end{equation}
Therefore $F(t)$ can be explicitly expressed by summing the marginal averages as
\begin{equation}
	\label{eq:decfunRTNqua}
	F(t)=e^{ic\nu^{2}t}.
\end{equation}
It is worth noting that the nonstationary statistics of the RTN only influences the marginal average $F(\pm\nu,t)$ but does not influence $F(t)$ due to its symmetric contributions in $F(\nu,t)$ and $F(-\nu,t)$.
The quantum system does not undergo decoherence in dynamical evolution since $\Gamma(t)=0$ for any time $t$.
Consequently, the frequency shift $S(t)$ can be expressed as 
\begin{equation}
	\label{eq:freshiRTNqua}
	\begin{split}
		S(t)=-c\nu^{2},
	\end{split}
\end{equation}
which is independent of both the nonstationary parameter $a$ and time $t$. 
This is due to the fact that in the case of the fluctuation term under quadratic dependence of RTN, the stochasticity of the fluctuation term $\delta\omega(t)$ vanishes which does not destruct the quantum coherence of the system but gives rise to a determinate frequency renormalization of the quantum system. 

\subsubsection{Comparison between the cases under the linear and nonlinear dependence of RTN}

Making a comparison of the results that the frequency difference fluctuates  linearly and quadratically in terms of the RTN, we can see that the environmental noise leads to dynamical decoherence for the linear case whereas it does not contribute to decoherence but causes the additional unitary contribution in dynamical evolution for the quadratic case.
This is also the significant difference in the quantum decoherence dynamics of the system in the presence of linear and quadratic environmental fluctuations governed by RTN.

\section{Conclusions}
\label{sec:con}
We have theoretically studied the quantum decoherence of a two-level system interacting with fluctuating environments by means of the approach of stochastic Liouville equation.
It is shown that there is a renormalization of the intrinsic energy levels of the quantum system in the case of either linear or quadratic dependence of the environmental noise.
In the former case, the frequency renormalization arises from the nonstationary statistical property of the environmental noise and vanishes in the case of environmental stationary statistics, whereas in the latter case, it always emerges even when the environmental noise exhibits stationary statistics.
In terms of the analytical expressions of the decoherence function, decoherence rate and frequency shift, we have performed detailed analyses of the nonlinear dependence of environmental fluctuations and the nonstationary statistics of the OUN and RTN on the dynamical decoherence of the quantum system, respectively.
For the case under the linear influence environmental noise, the quantum decoherence of the system is enhanced by the nonstationary statistical property of the OUN whereas it is suppressed by the nonstationary statistics of the RTN.
In the case of the nonlinear dependence of the environmental noise, the nonstationary statistical property of the OUN can reduce the dynamical decoherence whereas the quantum system does not undergo decoherence in dynamical evolution under the quadratic influence of the RTN.
We hope that the analytical solutions of the physical models and the analysis of the environmental effects in the study can help to further understand the dynamical decoherence of open quantum systems and can have the potential application in the design and realization of quantum devices with long coherent time in experiments.

\begin{acknowledgments}
	This work was supported by the National Natural Science Foundation of China  (Grant Nos. 12005121 and 12174221) and the Natural Science Foundation of Shandong Province (Grant No. ZR2021LLZ009).
\end{acknowledgments}

\section*{Author declarations}
\subsection*{Conflict of Interest}
The authors have no conflicts to disclose.
\subsection*{Author Contributions}
\textbf{Xiangji Cai:} Conceptualization (equal); Formal analysis (equal); Investigation (equal); Writing – original draft (lead); Writing – review \& editing (equal). 
\textbf{Yanyan Feng:} Conceptualization (equal); Formal analysis (equal); Investigation (equal); Writing – review \& editing (equal). 
\textbf{Jin Ren:} Conceptualization (equal); Formal analysis (equal); Investigation (equal); Writing – review \& editing (equal). 
\textbf{Yonggang Peng:} Conceptualization (equal); Formal analysis (equal); Investigation (equal); Writing – review \& editing (equal).
\textbf{Yujun Zheng:} Conceptualization (equal); Formal analysis (equal); Investigation (equal); Supervision (lead); Writing – review \& editing (equal).

\section*{Data Availability}
The data that support the findings of this study are available from the corresponding author upon reasonable request.

\appendix

\section{Quantum dephasing of a $d$-level system induced by classical environmental fluctuations}
\label{sec:App0}
Based on the framework established by Anderson and Kubo, the environmental effects of the classical noisy environments on a $d$-level quantum system phenomenologically leads to the stochastic fluctuations in the Hamiltonian of the system as~\cite{JPhysSocJpn.9.316,JPhysSocJpn.9.935}
\begin{equation}
	\label{eq:stomodApp0}
	H(t)=H_{0}+\delta H(t)=\hbar\sum_{n}[\omega_{n}+\delta\omega_{n}(t)]|n\rangle\langle n|,
\end{equation}
where $|n\rangle$ denotes a set of orthonormal eigenstates of the intrinsic Hamiltonian $H_{0}$ with unperturbed energies $E_{n}=\hbar\omega_{n}$, and the eigenenergies of the system fluctuates $E_{n}(t)=\hbar[\omega_{n}+\delta\omega_{n}(t)]$ with the environmental stochastic modulation $\delta\omega_{n}(t)$.
	
In the presence of stochastic environmental fluctuations, the density matrix of the quantum system evolves stochastically in time. The dynamical evolution of the stochastic density matrix is governed by the Liouville equation
\begin{equation}
	\label{eq:dynevoApp0}
	\frac{\partial}{\partial t}\rho\bm(t;\delta H(t)\bm)=-\frac{i}{\hbar}[H(t),\rho\bm(t;\delta H(t)\bm)],
\end{equation}
where $\rho\bm(t;\delta H(t)\bm)$ is adopted to indicate that the system dynamics depends on the stochastic Hamiltonian $\delta H(t)$.
The quantum system undergoes pure dephasing due to the fact that the stochastic  Hamiltonian induced by the environmental fluctuations is compatible with  the intrinsic Hamiltonian of the system, namely, $[H_{0},\delta H(t)]=0$.

The reduced density matrix of the quantum system can be, by averaging over different realizations of the environmental fluctuations, expressed as
\begin{equation}
	\label{eq:reddenmatApp0}
	\begin{split}
		\rho(t)&=\langle\rho(t;\delta H(t))\rangle\\
		&=\sum_{m,n}\rho_{mn}(0)e^{-i(\omega_{m}-\omega_{n})t}F_{mn}(t)|m\rangle\langle n|,
	\end{split}
\end{equation}
where $F_{mn}(t)=\big\langle\exp[-i\int_{0}^{t}\delta\omega_{mn}(t')dt']\big\rangle$ is the decoherence function describing the loss of coherence between states $|m\rangle$ and $|n\rangle$ with the stochastic frequency difference $\delta\omega_{mn}(t)=\delta\omega_{m}(t)-\delta\omega_{n}(t)$ which is a function of a Markovian stochastic process.

In the presence of the classical dephasing environment, the off-diagonal elements of the reduced density matrix of the quantum system in Eq.~\eqref{eq:reddenmatApp0} change with time while the diagonal elements are time independent.	
The decoherence function $F_{mn}(t)$ can be derived based on the theoretical framework of  stochastic Liouville equation established in Sec.~\ref{sec:theo}.
Once the decoherence function $F_{mn}(t)$  between states $|m\rangle$ and $|n\rangle$  is obtained, we can easily obtain some other information related to the dynamics of the quantum system, such as, the decoherence rate $\Gamma_{mn}(t)$ and the frequency shift $S_{mn}(t)$.

The two-level approximation for a quantum system has been widely used in nuclear magnetic resonance and quantum optics. This approximation is highly valid in cases where the transition energy between the lowest two levels differs from the transition energies to other levels.
A two-level system is also known as a two state system or a qubit, such as a nucleus of spin $1/2$ or a photon with two polarization states, which is a fundamental and important physical model in quantum information science~\cite{RevModPhys.89.015001}.
For the case of a two-level quantum system coupled to a classical dephasing environment, its Hamiltonian fluctuates stochastically in the form of Eq.~\eqref{eq:stoHam}.

\section{Stochastic processes with nonstationary statistical properties}
\label{sec:AppA}

In this appendix, we give the descriptions of the statistical properties of nonstationary stochastic processes based on the classical probability theory.
The influences of a fluctuating environment on an open quantum system can be generally described by the environmental noise which is subject to a Markovian stochastic process. 
The statistical properties of the environmental noise characterize the state of the environment and its interaction with the system.

For a Markovian stochastic process $x(t)$, the transition probability $P(x,t|x',t')$ that takes the value $x$ at time $t$ given that it took the value $x'$ at time $t'$ ($t'<t$) fulfills a master equation as Eq.~\eqref{eq:trapro} and satisfies the relations
\begin{equation}
	\label{eq:traprorelAppA}
	\begin{split}
		\int P(x,t|x',t') dx&=1,\\
		\lim_{t\rightarrow t'}P(x,t|x',t')&=\delta(x-x').
	\end{split}
\end{equation}

Based on the Markovian statistical property, the $n$-point probability of the stochastic process $x(t)$ can be expressed as
\begin{equation}
	\label{eq:MarconAppA}
	P(x_{1},t_{1};\cdots; x_{n},t_{n})=\prod_{i}^{n-1}P(x_{i+1},t_{i+1}|x_{i},t_{i})P(x_{1},t),
\end{equation}
and the single-point probability $P(x,t )$ that the stochastic process $x(t)$ has the value $x$ at time $t$ is connected to the initial probability $P(x_{0},0)$ at time $t_{0}=0$ with the relation
\begin{equation}
	\label{eq:sinpoiproAppA}
	P(x,t)=\int P(x,t|x_{0},0)P(x_{0},0)dx_{0}.
\end{equation}
The initial single-point probability $P(x,0)$ and the transition probability $P(x,t|x',t')$ fully determine the whole hierarchy of the multi-point probabilities $P(x_{1},t_{1};\cdots;x_{n},t_{n})$ of the process $x(t)$.

The Markovian stationary stochastic processes are widely used to describe equilibrium environmental fluctuations in both the statistical and quantum  physics~\cite{Kampenbook}.
The stochastic process $x(t)$ is stationary if the whole hierarchy of the multi-point probabilities depends on the time differences alone, namely, invariant under time translations for all $\tau$ 
\begin{equation}
	\label{eq:MarstaconAppA}
	P(x_{1},t_{1}+\tau;\cdots;x_{n},t_{n}+\tau)=P(x_{1},t_{1};\cdots;x_{n},t_{n}).
\end{equation}
In particular, the stationary statistical property of the process $x(t)$ implies that the single-point probability $P(x,t)=P(x)$ is independent of time corresponding to the single-point probability $P(x,0)$ in Eq.~\eqref{eq:sinpoiproAppA} with a stationary distribution initially and the transition probability $P(x,t|x',t')$ depends only on the time difference $t-t'$ of its time arguments.
Thus, the stochastic process $x(t)$ is homogeneous in time if it is stationary.

The Markovian stochastic processes with nonstationary statistical properties have wide applications in the description of nonequilibrium environmental fluctuations~\cite{Kampenbook}. 
The whole hierarchy of the multi-point probabilities of a time-homogeneous nonstationary process can be actually extracted from that of a Markovian stationary one. 
by setting the initial single-point probability $P(x,0)$ in Eq.~\eqref{eq:sinpoiproAppA} with a nonstationary distribution and the transition probability $P(x,t|x',t')$ satisfying the master equation~\eqref{eq:trapro}.
On the basis of the nonstationary distribution of the environmental noise, it implies physically that the state of the environment is nonstationary initially and it will return to stationary in long time limit. 

Based on the extraction of statistical property of a nonstationary process from that of a stationary one established above, we can choose the single-point probability distribution
in Eq.~\eqref{eq:inisinpoiOUN} for nonstationary OUN and set the single-point probability distribution initially in Eq.~\eqref{eq:inisinpoiRTN} for nonstationary RTN.
It is worth noting that both the odd-order moments and cumulants vanish if the OUN and RTN exhibit stationary statistics with zero average due to the time-homogeneous property.

\section{Solutions of the marginal average $F(x,t)$ under the linear and quadratic influences of nonstationary OUN}
\label{sec:AppB}
In this appendix, we give the derivations of the analytical expressions of the marginal average $F(x,t)$ for the decoherence function in the linear and quadratic dependence of the nonstationary OUN in Eqs.~\eqref{eq:comaveOUNlin} and~\eqref{eq:comaveOUNqua}, respectively.

To derive the marginal average $F(x,t)$ in Eq.~\eqref{eq:maraveOUN} analytically, 
we first make a transformation as
\begin{equation}
	\label{eq:traAppB}
	\bar{F}(x,t)=e^{\frac{x^{2}}{4\sigma^{2}}}F(x,t),
\end{equation}
which is governed by the time evolution
\begin{equation}
	\label{eq:tramaraveOUNAppB}
	\frac{\partial}{\partial t}\bar{F}(x,t)=\mathbb{M}\bar{F}(x,t),
\end{equation}
with the marginal average operator 
\begin{equation}
	\label{eq:maraveopeAppB}
	\mathbb{M}=\gamma\sigma^{2}\frac{\partial^{2}}{\partial x^{2}}-\frac{\gamma}{4\sigma^{2}}x^{2}+icx^{k}+\frac{\gamma}{2}.
\end{equation}
%
%we rewrite 
%its time evolution as
%\begin{equation}
%	\label{eq:maraveOUNAppB}
%	\frac{\partial}{\partial t}F(x,t)=\mathbb{M}F(x,t),
%\end{equation}
%where the marginal average operator $\mathbb{M}=icx^{k}+\gamma(\partial/\partial x)\left[x+\sigma^{2}(\partial/\partial x)\right]$ is non-Hermitian. 

By means of the approach of eigenfunction expansion,  we can express the solution of $\bar{F}(x,t)$ in the form~\cite{PhysRevA105.052443}
\begin{equation}
	\label{eq:eigexpAppB}
	\bar{F}(x,t)=\sum_{n}c_{n}(0)e^{\lambda_{n}t}\psi_{n}(x).
\end{equation}
Here $\psi_{n}(x)$ are the right eigenfunctions of the marginal average operator $\mathbb{M}$ in Eq.~\eqref{eq:maraveopeAppB} with the eigenvalues $\lambda_{n}$, and the initial expansion coefficients $c_{n}(0)$ can be expressed as
\begin{equation}
	\label{eq:eigexpAppB}
	c_{n}(0)=\int\tilde{\psi}_{n}(x_{0})\bar{F}(x_{0},0)dx_{0},
\end{equation}
where $\tilde{\psi}_{n}(x)$ denote the left eigenfunctions of the operator $\mathbb{M}$ and they are the Hermitian conjugate of the right eigenfunctions $\psi_{n}(x)$ when $\mathbb{M}$  is a Hermitian operator. %orthogonality condition
The left and right eigenfunctions satisfy the orthonormality relation $\int\tilde{\psi}_{n}(x)\psi_{n'}(x)dx=\delta_{nn'}$~\cite{Moiseyevbook,PhysRevA86.064104}.
In the cases of $k=1$ and $k=2$, the eigenvalues $\lambda_{n}$ and the eigenfunctions $\psi_{n}(x)$ and $\tilde{\psi}_{n}(x)$ of the operator $\mathbb{M}$ can be derived within the framework of quantum mechanics.

Thus, the marginal average $F(x,t)$ in Eq.~\eqref{eq:maraveOUN} can be formally expressed as
\begin{equation}
	\label{eq:maraveOUNAppB}
	\begin{split}
		F(x,t)&=e^{-\frac{x^{2}}{4\sigma^{2}}}\bar{F}(x,t)\\
		&=\sum_{n}\int\tilde{\psi}_{n}(x_{0})e^{\frac{x_{0}^{2}}{4\sigma^{2}}}F(x_{0},0)dx_{0}e^{\lambda_{n}t}\psi_{n}(x)e^{-\frac{x^{2}}{4\sigma^{2}}},
	\end{split}
\end{equation}
where $F(x_{0},0)=e^{-\frac{(x_{0}-b\chi)^{2}}{2\sigma^{2}(1-b^{2})}}/\sqrt{2\pi\sigma^{2}(1-b^{2})}$ is the initial value.

\subsection{Marginal average $F(x,t)$ in the linear dependence of nonstationary OUN}
\label{AppB1}
For the case under the linear influence of nonstationary OUN, namely, $k=1$, the marginal average operator in Eq.~\eqref{eq:maraveopeAppB} can be further simplified as
\begin{equation}
	\label{eq:modmaraveopelinAppB}
	\mathbb{M}=\gamma\sigma^{2}\frac{\partial^{2}}{\partial x^{2}}-\frac{\gamma}{4\sigma^{2}}\left(x-\frac{2ic\sigma^{2}}{\gamma}\right)^{2}+\frac{\gamma}{2}-\frac{c^{2}\sigma^{2}}{\gamma}.
\end{equation}
Correspondingly, the eigenvalues and eigenfunctions of the operator $\mathbb{M}$ can be expressed as
\begin{equation}
	\label{eq:modlineigAppB}
	\begin{split}
		\lambda_{n}&=-n\gamma-\frac{c^{2}\sigma^{2}}{\gamma},\\
		\psi_{n}(x)&=\tilde{\psi}_{n}(x)=\left(\frac{1}{2\pi\sigma^{2}}\right)^{1/4}\frac{1}{\sqrt{2^{n}n!}}H_{n}\left(\frac{x-2ic\sigma^{2}/\gamma}{\sqrt{2\sigma^{2}}}\right)\\
		&\qquad\qquad\quad\times\exp\left[-\frac{\left(x-2ic\sigma^{2}/\gamma\right)^{2}}{4\sigma^{2}}\right],
	\end{split}
\end{equation}
where $H_{n}(y)$ denotes the $n$th-order Hermite polynomial of the variable $y$, and the integer $n=0,1,2,\cdots$.

The marginal average $F(x,t)$ for the liner case can be written as 
\begin{equation}
	\label{eq:comaveOUNlinAppB}
	\begin{split}
		F(x,t)&=\frac{\exp\left(-c^{2}\sigma^{2}t/\gamma\right)}{2\pi\sigma^{2}\sqrt{(1-b^{2})}}\int\exp\bigg[-\frac{(x_{0}-b\chi)^{2}}{2\sigma^{2}(1-b^{2})}\\
		&\quad-\frac{\left(x-2ic\sigma^{2}/\gamma\right)^{2}}{2\sigma^{2}}-\frac{ic(x_{0}+x)}{\gamma}\bigg]\sum_{n=0}^{\infty}\frac{e^{-n\gamma t}}{2^{n}n!}\\
		&\quad\times H_{n}\left(\frac{x_{0}-2ic\sigma^{2}/\gamma}{\sqrt{2\sigma^{2}}}\right)H_{n}\left(\frac{x-2ic\sigma^{2}/\gamma}{\sqrt{2\sigma^{2}}}\right)dx_{0}.
	\end{split}
\end{equation}
Based on Eq.~\eqref{eq:comaveOUNlinAppB} and Mehler's summation formula for the Hermite polynomials ($|\zeta|<1$)~\cite{Szegobook}
\begin{equation}
	\label{eq:sumforAppB}
	\begin{split}
		\sum_{n=0}^{\infty}\frac{\zeta^{n}}{2^{n}n!}H_{n}(y)H_{n}(y')&=\frac{1}{\sqrt{1-\zeta^{2}}}\\
		&\quad\times\exp\left[\frac{2\zeta yy'-\zeta^{2}(y^{2}+y'^{2})}{1-\zeta^{2}}\right],
	\end{split}
\end{equation}
we can obtain the marginal average $F(x,t)$ for the case in the linear dependence of nonstationary OUN expressed in Eq.~\eqref{eq:comaveOUNlin}.

\subsection{Marginal average $F(x,t)$ in the quadratic dependence of nonstationary OUN}
\label{AppB2}
For the case in the quadratic dependence of nonstationary OUN $(k=2)$, the marginal average operator can be further reduced to
\begin{equation}
	\label{eq:modmaraveopequaAppB}
	\mathbb{M}=\gamma\sigma^{2}\frac{\partial^{2}}{\partial x^{2}}-\left(\frac{\gamma}{4\sigma^{2}}-ic\right)x^{2}+\frac{\gamma}{2},
\end{equation}
The corresponding eigenvalues and eigenfunctions of the operator $\mathbb{M}$ can be expressed as
\begin{equation}
	\label{eq:modquaeigAppB}
	\begin{split}
		\lambda_{n}&=-n\epsilon\gamma-\frac{\gamma}{2}(\epsilon-1),\\
		\psi_{n}(x)&=\tilde{\psi}_{n}(x)=\left(\frac{\epsilon}{2\pi\sigma^{2}}\right)^{1/4}\frac{1}{\sqrt{2^{n}n!}}H_{n}\left(\sqrt{\frac{\epsilon}{2\sigma^{2}}}x\right)\\
		&\qquad\qquad\quad\times\exp\left(-\frac{\epsilon x^{2}}{4\sigma^{2}}\right),
	\end{split}
\end{equation}
where $\epsilon=\sqrt{1-4ic\sigma^{2}/\gamma}$. 

The marginal average $F(x,t)$ for the quadratic case can be expressed as 
\begin{equation}
	\label{eq:comaveOUNquaAppB}
	\begin{split}	
		F(x,t)&=\frac{\sqrt{\epsilon}\exp\left[-\gamma(\epsilon-1)t/2\right]}{2\pi\sigma^{2}\sqrt{(1-b^{2})}}\int\exp\bigg[-\frac{(x_{0}-b\chi)^{2}}{2\sigma^{2}(1-b^{2})}\\
		&\quad+\frac{(1-\epsilon)x_{0}^{2}-(1+\epsilon)x^{2}}{4\sigma^{2}}\bigg] \sum_{n=0}^{\infty}\frac{e^{-n\epsilon\gamma t}}{2^{n}n!}\\
		&\quad\times H_{n}\left(\sqrt{\frac{\epsilon}{2\sigma^{2}}}x_{0}\right)H_{n}\left(\sqrt{\frac{\epsilon}{2\sigma^{2}}}x\right)dx_{0}.
	\end{split}	
\end{equation}
On the basis of Eq.~\eqref{eq:comaveOUNquaAppB} and Mehler's summation formula in Eq.~\eqref{eq:sumforAppB}, we can obtain the solution of the marginal average $F(x,t)$ for the case under the quadratic influence of nonstationary OUN in Eq.~\eqref{eq:comaveOUNqua}.

\section{Solutions of the marginal averages $F(\pm\nu,t)$ in the linear and quadratic dependence of nonstationary RTN}
\label{sec:AppC}

In this appendix, we give the derivations of the analytical expressions of the marginal averages $F(\pm\nu,t)$ for the decoherence function under the linear and quadratic influences of the nonstationary RTN in Eqs.~\eqref{eq:maraveRTNlin} and~\eqref{eq:maraveRTNqua} by means of the Laplace transform technique, respectively.

By taking the Laplace transform over Eq.~\eqref{eq:maraveRTN},  we obtain two coupled linear equations
\begin{equation}
	\label{eq:maraveRTLPAppC}
	\begin{split}
		p\tilde{F}(\nu,p)-F(\nu,0)&=ic\nu^{k} \tilde{F}(\nu,p)-\lambda \tilde{F}(\nu,p)\\
		&\quad+\lambda \tilde{F}(-\nu,p),\\
		p\tilde{F}(-\nu,p)-F(-\nu,0)&=ic(-\nu)^{k} \tilde{F}(-\nu,p)-\lambda \tilde{F}(-\nu,p)\\
		&\quad+\lambda \tilde{F}(\nu,p),\\
	\end{split}
\end{equation}
where $\tilde{F}(\nu,p)=\int_{0}^{\infty}F(\pm\nu,t)e^{-pt}dt$.
The marginal averages for the decoherence function can be solved in Laplace domain as
\begin{equation}
	\label{eq:maraveRTLPsolAppC}
	\begin{split}
				\tilde{F}(\nu,p)&=\frac{\left[p+\lambda-ic(-\nu)^{k}\right]F(\nu,0)+\lambda F(-\nu,0)}{\left(p+\lambda-ic\nu^{k}\right)\left[p+\lambda-ic(-\nu)^{k}\right]-\lambda^{2}},\\
		\tilde{F}(-\nu,p)&=\frac{\lambda F(\nu,0)+\left(p+\lambda-ic\nu^{k}\right)F(-\nu,0)}{\left(p+\lambda-ic\nu^{k}\right)\left[p+\lambda-ic(-\nu)^{k}\right]-\lambda^{2}},
	\end{split}
\end{equation}
where the initial conditions satisfy $F(\pm\nu,0)=(1\pm a)/2$.

\subsection{Marginal averages $F(\pm\nu,t)$ under the linear influence of nonstationary RTN}
\label{AppC1}
For the case in the linear dependence of nonstationary RTN $(k=1)$, in Laplace domain the marginal averages in Eq.~\eqref{eq:maraveRTLPAppC} can be reduced to
\begin{equation}
	\label{eq:maraveRTLP2AppC}
	\begin{split}
		\tilde{F}(\nu,p)&=\frac{(1+a)(p+\lambda+ic\nu)+(1-a)\lambda}{2\left[p(p+2\lambda)+(c\nu)^{2}\right]}\\
		&=\frac{(1+a)p+2\lambda+i(1+a)c\nu}{2\left[p(p+2\lambda)+(c\nu)^{2}\right]},\\\\
		\tilde{F}(-\nu,p)&=\frac{(1+a)\lambda+(1-a)(p+\lambda-ic\nu)}{2\left[p(p+2\lambda)+(c\nu)^{2}\right]}\\
		&=\frac{(1-a)p+2\lambda-i(1-a)c\nu}{2\left[p(p+2\lambda)+(c\nu)^{2}\right]},\\
	\end{split}
\end{equation}
By taking the inverse Laplace transform over Eq.~\eqref{eq:maraveRTLP2AppC}, we obtain the time dependent marginal averages for the decoherence function
\begin{equation}
	\label{eq:maraveRTNtimAppC}
	\begin{split}	
		F(\nu,t)=e^{-\lambda t}\left[\lambda m(t)+\frac{1}{2}(1+ a)[n(t)+ic\nu m(t)]\right],\\
		F(-\nu,t)=e^{-\lambda t}\left[\lambda m(t)+\frac{1}{2}(1-a)[n(t)-ic\nu m(t)]\right],\\
	\end{split}
\end{equation}
where the auxiliary functions $m(t)=(1/\kappa)\sinh(\kappa t)$ and $n(t)=\cosh(\kappa t)-(\lambda/\kappa)\sinh(\kappa t)$ with $\kappa=\sqrt{\lambda^{2}-(c\nu)^{2}}$.
This is the expression of the marginal averages $F(\pm\nu,t)$ for the case in the linear dependence of nonstationary RTN derived in Eq.~\eqref{eq:maraveRTNlin}.

\subsection{Marginal averages $F(\pm\nu,t)$ under the quadratic influence of nonstationary RTN}
\label{AppC2}
For the case under the quadratic influence of nonstationary RTN, namely, $k=2$, the marginal averages for the decoherence function in Eq.~\eqref{eq:maraveRTLPAppC} in Laplace domain can be simplified as
 \begin{equation}
	\label{eq:maraveRTLP3AppC}
	\begin{split}
		\tilde{F}(\nu,p)
		&=\frac{(1+a)(p-ic\nu^{2}+\lambda)+(1-a)\lambda}{2\left[(p-ic\nu^{2})(p-ic\nu^{2}+2\lambda)\right]}\\
		&=\frac{(1+a)(p-ic\nu^{2})+2\lambda}{2\left[(p-ic\nu^{2})(p-ic\nu^{2}+2\lambda)\right]},\\
		\tilde{F}(-\nu,p)
		&=\frac{(1+a)\lambda+(1-a)(p-ic\nu^{2}+\lambda)}{2\left[(p-ic\nu^{2})(p-ic\nu^{2}+2\lambda)\right]}\\
		&=\frac{(1-a)(p-ic\nu^{2})+2\lambda}{2\left[(p-ic\nu^{2})(p-ic\nu^{2}+2\lambda)\right]}.
	\end{split}
\end{equation}
By means of the inverse Laplace transform taken over Eq.~\eqref{eq:maraveRTLP3AppC}, we obtain the time evolution of the marginal averages for the decoherence function as
\begin{equation}
	\label{eq:maraveRTNquaAppC}
	\begin{split}
		F(\pm\nu,t)&=\frac{1}{2}\left[(1\pm a)e^{-2\lambda t}+(1-e^{-2\lambda t})\right]e^{ic\nu^{2}t}\\
		&=\frac{1}{2}\left(1\pm ae^{-2\lambda t}\right)e^{ic\nu^{2}t}.
	\end{split}
\end{equation}
This is the derivation of the marginal averages $F(\pm\nu,t)$ for the case under the quadratic influence of nonstationary RTN expressed in Eq.~\eqref{eq:maraveRTNqua}.

\section{Closed dynamical equation for the decoherence function under linear influences of nonstationary OUN and RTN}
\label{sec:AppD}

In this appendix, we derive the analytical expressions of the decoherence function in the case of linear dependence of the environmental noise exhibiting nonstationary OUN and RTN statistical characteristics by means of closed dynamical equations.
As comparisons, we will show the consistency of the expressions of the decoherence function with those derived in Eqs.~\eqref{eq:decfunOUNlin} and~\eqref{eq:decfunRTNlin}. This demonstrates the validity of the approach of stochastic Liouville equation under the influence of the environmental noise with nonstationary statistics.

\subsection{Closed evolution equation for the decoherence function in the case of linear dependence of nonstationary OUN}
\label{sec:AppD.1}
The expression of the transition probability for the OUN in Eq.~\eqref{eq:traproOUN} can be solved as
\begin{equation}
	\label{eq:solconproOUNAppD}
	\begin{split}
		P(x,t|x',t')&=\frac{1}{\sqrt{2\pi\sigma^{2}\left(1-e^{-2\gamma(t-t')}\right)}}\\
		&\quad\times\exp\left\{-\frac{\left[x-x'e^{-\gamma(t-t')}\right]^{2}}{2\sigma^{2}\left[1-e^{-2\gamma(t-t')}\right]}\right\}.
	\end{split}
\end{equation}
In terms of Eqs.~\eqref{eq:inisinpoiOUN} and~\eqref{eq:solconproOUNAppD}, the single-point probability can be expressed as
\begin{equation}
	\label{eq:sinpoiOUNAppD}
	\begin{split}
		P(x,t)=&\int P(x,t|x_{0},0)P(x_{0},0)dx_{0}\\
		=&\frac{1}{\sqrt{2\pi\sigma^{2}(1-b^{2}e^{-2\gamma t})}}\exp\left[-\frac{(x-b\chi e^{-\gamma t})^{2}}{2\sigma^{2}(1-b^{2}e^{-2\gamma t})}\right].
	\end{split}
\end{equation}

We can, based on the statistical properties of the nonstationary OUN $x(t)$ in terms of  Eqs.~\eqref{eq:solconproOUNAppD} and~\eqref{eq:sinpoiOUNAppD} above, 
derive its average as
\begin{equation}
	\label{eq:aveOUNAppD}
	\langle x(t)\rangle=\int xP(x,t)dx=b\chi e^{-\gamma t},
\end{equation}
and the second-order moment as
\begin{equation}
	\label{eq:corfunOUNAppD}
	\begin{split}
		\langle x(t)x(t')\rangle&=\int xx'P(x,t|x,t')P(x',t')dxdx'\\
		&=e^{-\gamma(t-t')}\left[\sigma^{2}(1-b^{2}e^{-2\gamma t'})+b^{2}\chi^{2}e^{-2\gamma t'}\right].
	\end{split}
\end{equation}
Thus, the time-order cumulant of second-order can be expressed as
\begin{equation}
	\label{eq:comOUNAppD}
	\begin{split}
		\langle x(t)x(t')\rangle_{\mathrm{oc}}&=\langle x(t)x(t')\rangle-\langle x(t)\rangle\langle x(t')\rangle\\
		&=\sigma^{2}\left[e^{-\gamma(t-t')}-b^{2}e^{-\gamma (t+t')}\right],
	\end{split}
\end{equation}
and the time-order cumulants higher than second order vanish based on the Markovian statistical property in Eq.~\eqref{eq:MarconAppA} can be written as
\begin{equation}
	\label{eq:higordcumAppD}
	\langle x(t)x(t_{1})\cdots x(t_{n})\rangle_{\mathrm{oc}}=0,
\end{equation}
for every set of time instants $t>t_{1}>\cdots>t_{n} (n\geq2)$.

Consequently, the dynamical evolution for the decoherence function in Eq.~\eqref{eq:dyson} is governed by a time-convolutionless equation closed at second-order as
\begin{equation}
	\label{eq:TCLmasequOUNAppD}
	\begin{split}
		\frac{d}{dt}F(t)&=ic\langle x(t)\rangle F(t)-c^{2}\int_{0}^{t}\langle x(t)x(t')\rangle_{\mathrm{oc}}F(t)dt'.
	\end{split}
\end{equation}
By solving Eq.~\eqref{eq:TCLmasequOUNAppD}, we obtain the analytical expression of the decoherence function under the linear influence of the nonstationary OUN as
\begin{equation}
	\label{eq:decfunOUTCLAppD}
	\begin{split}
		F(t)&=\exp\left\{i\frac{cb\chi}{\gamma}\eta(t)-\frac{(c\sigma)^{2}}{\gamma^{2}}\left[\gamma t-\eta(t)+\frac{1}{2}b^{2}\eta^{2}(t)\right]\right\},
	\end{split}
\end{equation}
with $\eta(t)=1-e^{-\gamma t}$.
The expression of $F(t)$ consists with that derived analytically by means of the approach of stochastic Liouville equation in Eq.~\eqref{eq:decfunOUNlin}.

\subsection{Closed time evolution equation for the decoherence function under  linear influence of nonstationary RTN}
\label{sec:AppD.2}

The solution of the transition probability of the RTN in Eq.~\eqref{eq:traproRTN} can be written as
\begin{equation}
	\label{eq:solconproRTNAppD}
	\begin{split}	
		P(x,t|x',t')&=\frac{1}{2}\left[1+e^{-2\lambda(t-t')}\right]\delta_{x,x'}\\
		&\quad+\frac{1}{2}\left[1-e^{-2\lambda(t-t')}\right]\delta_{x,-x'}.
	\end{split}
\end{equation}
Based on Eqs.~\eqref{eq:inisinpoiRTN} and~\eqref{eq:solconproRTNAppD}, the single-point probability can be expressed as
\begin{equation}
	\label{eq:sinpoiRTNAppD}
	\begin{split}
		P(x,t)&=\sum_{x_{0}} P(x,t|x_{0},0)P(x_{0},0)\\
		&=\frac{1}{2}(1+ae^{-2\lambda t})\delta_{x,\nu}+\frac{1}{2}(1-ae^{-2\lambda t})\delta_{x,-\nu}.
	\end{split}
\end{equation}

According to the statistical properties in terms of Eqs.~\eqref{eq:solconproRTNAppD} and~\eqref{eq:sinpoiRTNAppD}, the average of the nonstationary RTN can be expressed as
\begin{equation}
	\label{eq:aveRTNAppD}
	\langle x(t)\rangle=\sum_{x} xP(x,t)=ac\nu e^{-2\lambda t},
\end{equation}
and its second-order moment can be written as
\begin{equation}
	\label{eq:corRTNAppD}
	\begin{split}
		\langle x(t)x(t')\rangle&=\sum_{x,x'}xx'P(x,t|x,t')P(x',t')\\
		&=\nu^{2}e^{-2\lambda(t-t')}.
	\end{split}
\end{equation}
Based on the Markovian statistical property in Eq.~\eqref{eq:MarconAppA}, the moments higher than second-order satisfy the factorization relation
\begin{equation}
	\label{eq:higmomAppD}
	\langle x(t)x(t_{1})\cdots x(t_{n})\rangle=\langle x(t)x(t_{1})\rangle\langle x(t_{2})\cdots x(t_{n})\rangle,
\end{equation}
for all sets of the time sequences $t>t_{1}>\cdots>t_{n} (n\geq2)$.

By taking the differentiation of Eq.~\eqref{eq:dyson} with respect to time $t$, we derive a closed time-convolution equation with an inhomogeneous term for the decoherence function as
\begin{equation}
	\label{eq:TCmasequRTNAppD}
	\frac{d}{dt}F(t)=-c^{2}\int_{0}^{t}\langle x(t)x(t')\rangle F(t')dt'+ic\langle x(t)\rangle F(0).
\end{equation}
The memory kernel in Eq.~\eqref{eq:TCmasequRTNAppD} is related to the second-order moment of the RTN and the inhomogeneous term is associated with the average of the environmental noise.
By means of the Laplace transform technique, we can express the decoherence function in the case of linear dependence of the nonstationary RTN analytically as
\begin{equation}
	\label{eq:comavedecfunRTlinAppD}
	F(t)=e^{-\lambda t}\left[\cosh(\kappa t)+\frac{\lambda}{\kappa}\sinh(\kappa t)+i\frac{ac\nu}{\kappa}\sinh(\kappa t)\right].
\end{equation}
The analytical expression of $F(t)$ is consistent with that derived by means of the approach of stochastic Liouville equation in Eq.~\eqref{eq:decfunRTNlin}.

\end{CJK}

\bibliography{References}

\end{document}